\documentclass[preprint2]{aastex}

\slugcomment{Accepted for publication in the Special GALEX Ap.J.Suppl. Issue}

\shorttitle{Radial variations in late type galaxies}
\shortauthors{Boissier et al}

\begin{document}

\title{Radial variation of attenuation and star formation in the largest 
late-type disks observed with GALEX}

\author{Samuel Boissier\altaffilmark{1,2}, 
Armando Gil de Paz\altaffilmark{1,3}, 
Alessandro Boselli\altaffilmark{2}, 
Barry F. Madore\altaffilmark{1,4},
V\'eronique Buat\altaffilmark{2}, 
Luca Cortese\altaffilmark{2}, 
Denis Burgarella\altaffilmark{2},
Juan Carlos Mu\~{n}oz Mateos\altaffilmark{3}, 
Tom A. Barlow\altaffilmark{5}, 
Karl Forster\altaffilmark{5},
Peter G. Friedman\altaffilmark{5},
D. Christopher Martin\altaffilmark{5},
Patrick Morrissey\altaffilmark{5},
Susan G. Neff\altaffilmark{10},
David Schiminovich\altaffilmark{12},
Mark Seibert\altaffilmark{5},
Todd Small\altaffilmark{5},
Ted K. Wyder\altaffilmark{5},
Luciana Bianchi\altaffilmark{6},
Jose Donas\altaffilmark{2},
Timothy M. Heckman\altaffilmark{8},
Young-Wook Lee\altaffilmark{7},
Bruno Milliard\altaffilmark{2},
R. Michael Rich\altaffilmark{11},
Alex S. Szalay\altaffilmark{8},
Barry Y. Welsh\altaffilmark{9}, 
Sukyoung K. Yi \altaffilmark{7}
}

\altaffiltext{1}{Observatories of the Carnegie Institution of Washington,
813 Santa Barbara St., Pasadena, CA 91101, USA, boissier,agpaz,madore@ociw.edu}

\altaffiltext{2}{Laboratoire d'Astrophysique de Marseille, BP 8, Traverse
du Siphon, 13376 Marseille Cedex 12, France}

\altaffiltext{3}{Departamento de Astrof\'{\i}sica y CC. de la Atm\'osfera, 
Universidad Complutense de Madrid,
Avda. de la Complutense, s/n,
E-28040 Madrid, Spain}

\altaffiltext{4}{Infrared Processing and Analysis Center
California Institute of Technology Mail Code 100-22 770 South Wilson
Avenue Pasadena, CA 91125, USA}

\altaffiltext{5}{California Institute of Technology, MC 405-47, 1200 East
California Boulevard, Pasadena, CA 91125, USA}

\altaffiltext{6}{Center for Astrophysical Sciences, The Johns Hopkins
University, 3400 N. Charles St., Baltimore, MD 21218, USA}

\altaffiltext{7}{Center for Space Astrophysics, Yonsei University, Seoul
120-749, Korea}

\altaffiltext{8}{Department of Physics and Astronomy, The Johns Hopkins
University, Homewood Campus, Baltimore, MD 21218, USA}

\altaffiltext{9}{Space Sciences Laboratory, University of California at
Berkeley, 601 Campbell Hall, Berkeley, CA 94720, USA}

\altaffiltext{10}{Laboratory for Astronomy and Solar Physics, NASA Goddard
Space Flight Center, Greenbelt, MD 20771, USA}

\altaffiltext{11}{Department of Physics and Astronomy, University of
California, Los Angeles, CA 90095, USA}

\altaffiltext{12}{Department of Astronomy, Columbia University, New York, 
NY 10027, USA}

\begin{abstract}
For a sample of 43 nearby, late-type galaxies, we have
investigated the radial variation of both the current star formation
rate and the dust-induced UV light attenuation.
To do this we have cross-correlated IRAS images and GALEX observations
for each of these galaxies, and compiled observations of the gas (CO
and HI) and metal-abundance gradients found in the literature.
We find that attenuation correlates with metallicity.  We
then use the UV profiles, corrected for attenuation, to study several
variants of the Schmidt law and conclude that our results are
compatible with a simple law similar to the one of \citet{kenni98a},
extending smoothly to lower surface densities, but with considerable
scatter.
We do not detect an abrupt break in the UV light at the threshold
radius derived from H$\alpha$ data (at which the H$\alpha$ profile
shows a break and beyond which only a few HII regions are usually
found). We interpret the H$\alpha$ sudden break not as a change in the
star formation regime (as often suggested) but as the vanishingly small
number of ionizing stars corresponding to low levels of star formation.
\end{abstract}


\section{Introduction}

Interstellar dust affects our view of galaxies from the UV to 
the near-infrared. High-redshift galaxies are commonly studied in the
rest-frame UV where dust effects can be  extremely severe, and our
estimation of the ``cosmic'' star formation rate (SFR) 
\citep[e.g.][]{schimi05} crucially depends on the corrections applied
to account for the dust attenuation of starlight. 

Radiative transfer models suggest that the ratio of far-infrared to UV
radiation is a reliable measure of the UV attenuation
\citep{buat96,witt00,panuzzo03}, depending weakly upon the geometry of
stars and dust, the extinction law, or the nature of the underlying
stellar population.
Since this ratio is not always available, the slope of the UV spectrum
reddening with respect to local starbursts
\citep{calzetti94,meurer95,meurer99} has been commonly used as a
metric to estimate the amount of attenuation.
This allows one to compare the attenuation (as measured by
the far infrared/UV ratio) to the slope of the UV spectrum in a
diagnostic plot, the commonly-called ``IRX-$\beta$ relationship''.
Several recent works, mostly based on GALEX observations, cast doubt
on the IRX-$\beta$ relation that has been most commonly used during
the recent years \citep{buat05,cortese06,seibert05,gil06}.
Independent studies have come to similar conclusions, e.g. 
\citet{bell02} for normal spirals and \citet{bell02b} for
individual regions in the LMC.

These works suggests either that there is no relationship (or a very
noisy one) or that a relationship exists but differs from
the starbursts one, depending on the selection of the sample (infrared
vs UV or optically-selected).

The largest angular sized galaxies in the {\it GALEX Atlas of Nearby
Galaxies} \citep{gil06} allow us to revisit the IRX-$\beta$
relationship. Indeed, several tens of nearby galaxies observed by
GALEX are large enough to have been resolved even by IRAS. The
combination of the UV data of GALEX and infrared of IRAS opens up the
possibility of looking at the spatially resolved IRX-$\beta$
relationship across a relatively large sample of galaxies. Even if the
number of resolved objects in this study is smaller than the total
number of objects used in the global studies, each of our galaxies
provides a number of independent points along the radial profile.
Regions situated at the same radius within a galaxy 
are relatively homogeneous (for instance similar metallicities),
thus radial profiles are physically relevant. 

Another compelling reason to study attenuation\footnote{
``Attenuation'' is the loss of flux due to the presence of dust. It is
sometimes considered that ``extinction'' should be used only for the
dust properties and not the global effect of dust on the galaxy light,
also depending on e.g. the geometry of the system. ``Extinction'' is
nevertheless often used with the same meaning as ``attenuation''.}
in these galaxies is that, as our closest neighbors, they have been
extensively surveyed at other wavelengths, and in the published
literature there are metallicity gradient and gas density profiles,
etc. that allow us to go one step further and try to characterize how
the properties of dust might depend on these other physical
quantities.
We also produce star formation rate radial profiles derived from the
UV (and not H$\alpha$ data), corrected for the dust attenuation for
all our galaxies.  These profiles can be used to study how the star
formation rate depends on the gas density, {\it i.e.} the so-called
Schmidt law. We will be able to use these star formation rate profiles
to finally study the threshold
radius beyond which little (if any) star formation is
observed. Many studies suggest that the threshold radius corresponds
to the point where gas density becomes lower than a critical density
for star formation \citep[e.g.][]{martin01}.
It is important to bring as many constraints as possible to bear on
the complex problem of star formation and this paper brings its share,
by using UV GALEX data.
Indeed, while many other studies of the Schmidt law
\citep{kenni98a,boissier03,wongblitz} are based on H$\alpha$ data, the
star formation rate in this paper is derived from the UV, which
``averages'' the star formation on longer time-scales than
H$\alpha$. 
Recombination lines as H$\alpha$ are emitted only in presence of
a large ionizing flux, coming mostly from massive 
stars ($M > 10 M_{\odot}$) with lifetimes shorter
than 2 $\times 10^7$ years. On the contrary, the UV continuum
is emitted by stars with lifetimes up to $10^8$ years
\citep[a review of the various star formation indicators and their associated lifetimes is given in][]{kenni98b}.
Because of the scarcity of UV data before GALEX, little work had been
done in the UV on the star formation law. Some exceptions are, that of
\citet{buat89b},
\citet{bersier94} concerning respectively M81 and M51, or
\citet{buat89a} and \citet{donas87} for studies using sample of nearby
galaxies but these papers looked only at the integrated properties of
galaxies, and employed rather simple recipes for dust correction. The
radial variation of the UV-derived star formation rate, correctly
dust-corrected via the far-infrared to UV ratio, is the subject of
this paper. This is an important work as we now know that the UV also
shows us low levels of star formation
\citep[XUV galaxies, e.g.][]{gildepaz05,thilker05}, in regions with
such low dust content that the far-infrared emission is undetected
by IRAS, and that the UV is likely to be weakly affected
by extinction.

In section \ref{secdata}, we discuss how the data-sets were assembled
and prepared in order to undertake this study of the radial variation
of the attenuation, of the star formation, and of their relation to
other physical quantities in those same galaxies. In section
\ref{secext}, we present our study of the attenuation derived from
these data, and the consequences concerning the resulting star formation law
are presented in section \ref{secstarform}.

\clearpage
\begin{deluxetable}{l r r r r r r r r r r r }
\tablecaption{\label{tabrefs}Sample: basic parameters used and references}
\startdata
Name      &  E(B-V) &  T  &  D~~   &  PA  &  D25  &  d25 & Res. & \multicolumn{3}{l}{~~~~~~~~~~References} \\
          &  (mag)  &     & (Mpc)& (deg)&  (') & (') & ('') & [O/H]     & HI & CO & V(R)   \\
\object{IC 1613}  &   0.025 & 10      &          0.8   &   50  &  16  & 15  & 106    & d      & 24  &    & 24     \\
\object{IC 2574}  &   0.036 &  9      &          4.0   &   50  &  13  &  5  & 110    & e      & 25  &    & 25     \\
\object{M31}      &   0.062 &  3      &          0.8   &   35  & 190  & 60  & 108    & i      & 28  & 28 & 18, 27 \\
\object{M33}      &   0.042 &  6      &          0.8   &   23  &  71  & 42  & 108    & i      & 17  & 19 & 17     \\
\object{M51a}     &   0.035 &  4      &          8.4   &  -17  &  11  &  7  & 117    & i      & 5   & 30 & 40     \\
\object{M64}      &   0.041 &  2      &         17.0   &  -65  &  10  &  5  & 116    &        & 10  & 36 & 10     \\  
\object{M81}      &   0.080 &  2      &          3.6   &  -23  &  27  & 14  & 111    & i      & 5   & 5  & 40     \\
\object{M83}      &   0.066 &  5      &          4.5   &    0  &  13  & 12  & 106    & i      & 6   & 6  & 6      \\
\object{M101}     &   0.009 &  6      &          7.5   &   90  &  29  & 27  &  88    & b      & 23  & 23 & 23     \\
\object{NGC 0055} &   0.013 &  9      &          2.0   &  -72  &  32  &  6  &  96    & i      & 33  &    & 33     \\
\object{NGC 0247} &   0.018 &  7      &          3.1   &   -6  &  21  &  7  & 107    &        & 14  &    & 14     \\
\object{NGC 0253} &   0.019 &  5      &          3.9   &   52  &  28  &  7  & 114    & i      & 34  & 41 & 34     \\
\object{NGC 0300} &   0.013 &  7      &          2.0   &  -69  &  22  & 16  & 106    & i      & 32  &    & 32     \\
\object{NGC 0628} &   0.070 &  5      &         11.2   &   25  &  11  & 10  & 111    &a, h, i & 39  & 36 & 39     \\
\object{NGC 0660} &   0.065 &  1      &         12.4   &  -10  &   8  &  3  & 108    &        &     & 46 & 43     \\
\object{NGC 0891} &   0.065 &  3      &          9.6   &   22  &  14  &  3  & 111    &        & 38  & 37 & 37, 38 \\
\object{NGC 0925} &   0.076 &  7      &          9.3   &  -78  &  11  &  6  & 110    & h, i   &  5  &  5 & 31     \\
\object{NGC 1097} &   0.027 &  3      &         15.3   &  -50  &   9  &  6  & 103    & g      &     & 46 & 29     \\
\object{NGC 1291} &   0.013 &  0      &          9.7   &  -15  &  10  &  8  &  88    &        & 42  &    &        \\
\object{NGC 1512} &   0.011 &  1      &         10.4   &   90  &   9  &  6  &  95    &        &     &    &        \\
\object{NGC 1566} &   0.009 &  4      &         17.5   &   60  &   8  &  7  &  95    & i      &     & 2  &        \\
\object{NGC 2366} &   0.036 & 10      &          3.4   &   25  &   8  &  3  &  95    & f      & 20  &    & 20     \\
\object{NGC 2403} &   0.040 &  6      &          3.2   &  -53  &  22  & 12  & 103    & h, i   & 5   & 5  & 3      \\
\object{NGC 2841} &   0.016 &  3      &         14.1   &  -33  &  08  &  4  & 105    &        & 3,9 & 45 & 3      \\
\object{NGC 2903} &   0.031 &  4      &          8.9   &   17  &  13  &  6  & 106    & h, i   & 5   & 5  & 3      \\
\object{NGC 3198} &   0.012 &  5      &         16.7   &   35  &   9  &  3  & 103    & i      & 4   &    & 8      \\
\object{NGC 3521} &   0.058 &  4      &          9.0   &  -17  &  11  &  5  & 104    & i      &     & 36 & 15     \\
\object{NGC 3621} &   0.080 &  7      &          8.3   &  -21  &  12  &  7  & 105    & i      &     &    &        \\
\object{NGC 3627} &   0.032 &  3      &          9.1   &  -07  &   9  &  4  & 110    &        & 47  & 36 & 47     \\
\object{NGC 4216} &   0.032 &  3      &         17.0   &   19  &   8  &  2  & 108    &        & 16  & 22 &        \\
\object{NGC 4236} &   0.015 &  8      &          4.5   &  -18  &  22  &  7  & 102    &        & 11  &    &        \\
\object{NGC 4395} &   0.017 &  9      &          4.2   &  -33  &  13  & 11  & 104    & h, f   & 44  &    & 44     \\
\object{NGC 4438} &   0.028 &  0      &         17.0   &   27  &   9  &  3  & 106    &        &     &    &        \\
\object{NGC 4559} &   0.018 &  6      &         17.0   &  -30  &  11  &  4  & 104    & i      & 12  &    & 12     \\
\object{NGC 4569} &   0.046 &  2      &         17.0   &   23  &  10  &  4  & 110    &        & 7   & 7  & 7      \\
\object{NGC 4594} &   0.051 &  1      &          9.1   &   90  &   9  &  4  & 108    &        & 1   &    & 1      \\
\object{NGC 4631} &   0.017 &  7      &          9.0   &   86  &  16  &  3  & 106    &        &     &    & 35     \\
\object{NGC 4656} &   0.013 &  9      &          9.0   &   33  &  15  &  3  & 106    &        &     &    &        \\
\object{NGC 4736} &   0.018 &  2      &          5.2   &  -75  &  11  &  9  & 104    & i      & 26  & 36 & 26     \\
\object{NGC 5055} &   0.018 &  4      &          8.2   &  -75  &  13  &  7  & 105    & i      & 9   & 36 & 9      \\
\object{NGC 7331} &   0.091 &  7      &         14.9   &  -09  &  11  &  4  & 104    & i      & 3   & 45 & 3      \\
\object{NGC 7793} &   0.019 &  7      &          2.0   &  -82  &   9  &  6  & 108    & i      & 13  &    & 13     \\
\object{WLM}      &   0.037 & 10      &          1.0   &  -20  &  12  &  4  & 104    & c      &     &    & 21     \\

\enddata
\tablecomments{The columns give 
(1) Object name, 
(2) Galactic color excess \citep{schlegel98} ,
(3) RC3 morphological type,
(4) Distance (Mpc) taken from the GALEX Atlas of nearby galaxies \citep{gil06},
(5) Position Angle (degree),
(6) D25: major axis diameter (minutes),
(7) d25: minor axis diameter (minutes),
(8) FWHM of the restored 100 microns beam in arcsec (see section \ref{secrestorebeam} for details),
(9) References for O/H gradients,
(10), (11), (12) References for HI, CO, and rotation curve data.
}
\tablerefs{ 
{\bf References for Abundance Gradients [O/H]:} 
a \citet{belley92} ;  
b \citet{kenni03a} ; 
c \citet{lee03aj} ; 
d \citet{lee03aa} ; 
e \citet{masegosa91} ; 
f \citet{roy96} ; 
g \citet{storchi96} ; 
h \citet{vanzee98} ; 
i \citet{zaritsky94}. 
{\bf References for Gas [HI, CO] and Kinematic [V(R)] Data:}     
1 \citet{bajaja84} ; 
2 \citet{bajaja95} ; 
3 \citet{begemanthesis} ;
4 \citet{begeman89} ; 
5 \citet{boissier03}, and references within ; 
6 \citet{boissier05}, and references within ; 
7 Boselli et al. (2006), and references within ;  
8 \citet{bosmathesis} ; 
9 \citet{bosma81} ; 
10 \citet{braun94} ; 
11 \citet{braun97} ; 
12 \citet{broeils94} ; 
13 \citet{carignan7793}; 
14 \citet{carignan247} ;
15 \citet{casertano91} ; 
16 \citet{cayatte94} ; 
17 \citet{corbelli03} ; 
18 \citet{dame93} ;
19 \citet{heyer04} ; 
20 \citet{hunter01} ; 
21 \citet{jackson04} ; 
22 \citet{kenney88} ; 
23 \citet{kenney91} ; 
24 \citet{lake89} ; 
25 \citet{martimbeau94} ; 
26 \citet{mulder93} 
27 \citet{newton77} ; 
28 \citet{nieten06} ; 
39 \citet{ondrechen89} ; 
30 \citet{paglione01} ; 
31 \citet{pisano98} ; 
32 \citet{puche90} ; 
33 \citet{puche9133} ; 
34 \citet{puche9134} ; 
35 \citet{rand94} ;        
36 \citet{regan01} ; 
37 \citet{sakamoto97} ; 
38 \citet{sancisi79} ; 
39 \citet{shostak84} ; 
40 \citet{sofue97} ; 
43 \citet{sorai00} ; 
44 \citet{vandriel88} ; 
45 \citet{vandriel95} ; 
46 \citet{wevers86} ; 
48 \citet{young82} ; 
50 \citet{young95} ; 
51 \citet{zhang93}.}  
\end{deluxetable}

\clearpage

\section{Profiles: Data and Procedure}
\label{secdata}

Our initial sample consists of 48 late-type galaxies (RC3 T type going
from 0 to +10, thus including all spirals from S0/a to Sm and
magellanic irregulars Im) with UV data in the GALEX Atlas of nearby
galaxies \citep{gil06}, larger than 8 arcmin (as measured by $D_{25}$
diameter corresponding to the B-band isophote of 25 mag
arcsec$^{-2}$.) This size allows for spatial resolution of these
galaxies at all the wavelengths needed, including the 60 and 100
$\mu$m IRAS data.

All the IRAS images were obtained using HIRES requests to the IPAC web
page\footnote{http://irsa.ipac.caltech.edu/IRASdocs/hires\_over.html}.
The method used in HIRES is described in \citet{aumann90}; examples of
images obtained with HIRES have been presented by \citet{rice93} for a
sample of nearby galaxies (The HIRES procedure failed to converge for
NGC~3109, and this galaxy was removed from our analysis).  
Because of the IRAS aperture itself and the complex pattern of
observations, the resulting HIRES images can be quite asymmetric.
%
HIRES produces simulated images of point sources distributed into the image
at the sky position of the targeted object.
The first step in our procedure was to apply an asymmetric Gaussian
convolution in order to produce a circular image at 60 $\mu$m and 100
$\mu$m of these point sources, and to apply the same convolution to the
HIRES reconstructed image of the galaxy.
\label{secrestorebeam}

In the UV, the GALEX FUV and NUV (Far and Near UV, respectively around
1516 and 2267 \AA) images of the Atlas \citep{gil06} were used. For
four galaxies, no FUV data were available (NGC~1365, NGC~3628, M98,
M106), and they are excluded from our analysis which \emph{in fine}
includes 43 galaxies.  Given this selection process, this sample is
not complete in any sense, although it should be representative of
optically selected galaxies \citep[a comparison of the
GALEX Atlas of nearby galaxies and the Nearby Field Galaxy 
Survey of Jansen et al. 2000 is given in][]{gil06}.

Stars were removed using the star masks of the Atlas and then
interpolated over. Foreground Galactic extinction was corrected as in
\citet{gil06} : using the values of the color excess E(B-V) given by
the maps of \citet{schlegel98} and a Galactic extinction curve.
\label{secGalext}

Images in FUV, NUV, and at 60 $\mu$m were then convolved with an
elliptical Gaussian function in order to match the resolution to the
one measured in the HIRES simulation of point-sources at 100 $\mu$m
(after the asymmetric Gaussian convolution discussed above).
The spatial resolution of
IRAS is relatively poor. However, this might help us to avoid
difficulties in interpreting the UV/IR balance since the dust of
small regions could be heated by the UV emission from neighboring
regions.

Profiles were finally computed within elliptical annuli with ELLIPSE in 
IRAF\footnote{
IRAF is distributed by the National Optical Astronomy Observatories,
which are operated by the Association of Universities for Research
in Astronomy, Inc., under cooperative agreement with the National 
Science Foundation.} 
using convolved images, with fixed center, ellipticity and position angles
(as given in Gil de Paz et al. 2006 and Table \ref{tabrefs}). The step
in major axis radius
between isophotes was chosen to be the measured resolution (close to
1.5 arcmin on average, and given in Table \ref{tabrefs}).

The infrared profiles were combined as in
\citet{dale01} in order to compute a ``far-infrared'' profile
($F_{FIR}$ in W m$^{-2}$ arcsec$^{-2}$) and a ``total-infrared'' profile ($F_{TIR}$), as described below: 
\begin{equation}
F_{FIR} = 1.26 \times 10^{-14}(2.58 f_{60} + f_{100})
\end{equation}
($f_{\lambda}$ are the IRAS flux surface densities in Jy arcsec$^{-2}$) and
\begin{eqnarray}
log(F_{TIR}) &=log(F_{FIR})  & +0.2738-0.0282 r+0.7281 r^2 \nonumber \\ 
   &           &  +0.6208 r^3+0.9118 r^4
\end{eqnarray}
where $r=log(f_{60}/f_{100})$

Attenuation profiles (expressed in magnitude) were then computed
using the fit of \citet{buat05}
\begin{eqnarray}
A(FUV) & = & -0.0333 X^3+0.3522 X^2 \nonumber \\
       &   &  +1.1960 X+0.4967 \label{eqbuatcal}
\end{eqnarray}
where $X=log (F_{TIR}/ F_{FUV})$ ; $F_{FUV}=\nu_{fuv} f_{\nu_{fuv}}$ and $f_{\nu_{fuv}}$ is the FUV
flux surface density in W m$^{-2}$ Hz$^{-1}$ arcsec$^{-2}$.  
Other expressions linking $A(FUV)$ and $X$ do exist
\citep[e.g.][]{burgarella05} with subtle differences, and 
since this kind of calibration might also be subject to future
revisions, we included in our figures both scales (observed $F_{TIR}/F_{FUV}$,
calibrated $A(FUV)$).

Given the long chain of operations in computing $A(FUV)$, we estimated
in a simple way its uncertainty by computing the extremal values it
can take when moving the UV and infrared fluxes by 1 $\sigma$ from
their observed values.
FIR and UV data are moreover uncertain by $\sim$ 15 \% on average,
(see e.g. Morrissey et al. 2005 for an assessment of the
quality of the GALEX photometry, and  Rice 1993 for 
the photometry of IRAS extended galaxies).

The additional uncertainty in the calibration of the relation 
given in equation \ref{eqbuatcal} itself is not taken into account.
Points and error-bars were not allowed to include negative
attenuations.  In the few cases where the $A(FUV)$ obtained in this
way was unphysical (e.g. for numerical reason), the last consistent
estimated error was used for points at larger radii (this only affects
a few outer points where error-bars are anyway large as can be seen in
the $A(FUV)$ profiles given in the appendix).\label{secerrorbar}

The FUV observed profiles were finally corrected for attenuation using
$A(FUV)$, corrected for inclination, and converted
to a star formation rate using the \citet{kenni98b} calibration.
%
The inclination correction accounts for the fact that the thickness of
a galaxy (and thus the amount of stars) along the line of sight has to
be multiplied by $cos(inc)$ ($inc$ being the inclination of the galaxy)  
to obtain the same values for a face-on 
orientation ($inc$=0). The adopted inclination is given by the ratio of 
$D25$ and $d25$ from Table 1, assuming the objects are intrinsically circular.
Note that this procedure assumes that we computed the correct $A(FUV)$
for the inclination of the galaxy. \citet{panuzzo03} showed that the
relation between $A(FUV)$ and the $F_{FIR}/F_{FUV}$ ratio actually depends on
inclination. Their fit for edge-on galaxies differs from the fit for
moderately inclined galaxies by up to about 0.3 magnitude of
extinction at 2000 \AA. There are other reasons why we should be
cautious about very inclined galaxies:
\label{secprobincli}
i) Our data points are separated by one element of resolution along
the major axis. In the case of inclined galaxies, the ellipses along
the minor axis will be closer to each other than the resolution. Outer
regions might therefore be contaminated by inner regions.
ii) In an edge-on disk, the amount of material along the line of sight
is such that only a full radiative-transfer modeling can be used to
reconstruct the UV to sub-millimeter spectral energy distribution of
galaxies \citep[e.g.][]{popescu2000}.
For this reason, our data are presented when possible with different
symbols for various axis ratio ranges, and we discussed in the text when the
inclination affected our results.

\label{secxco}
Corollary data were collected from the literature, concerning the
gaseous content (HI, CO, Rotation curve) and the oxygen abundance
gradient of the galaxies. CO data were converted to molecular H$_2$
using the conversion factor of \citet{boselli02xco}, dependent on the
metallicity when an abundance gradient was available.  For the seven
galaxies without metallicity gradient, we used the conversion factor
corresponding to the H-band luminosity of each galaxy, using the
calibration of \citet{boselli02xco}.

The gas profiles taken from the literature were converted to the same
inclination as mentioned before (indicated by the index ``inc''). We
also determined face on values (index ``0'') of the same quantities
performing a similar correction than the one applied to the UV surface
brightness (equivalent to a star formation rate surface density). Note
that face-on values should be used to compare star formation rate and
gas surface densities; while the extinction profiles obtained for the
observed galaxy inclination should be compared to the column densities
along the line of sight, i.e.  the ones with the ``inc'' index.
While we corrected the original gas data to have consistent
inclination with our own UV and far-infrared computations, we must
acknowledge that this compilation introduces some level of uncertainty
since all observations were not performed with the same instruments,
spatial resolution, etc. We however consider that this effect is included
within the typical uncertainties discussed below.

Typical uncertainties are 0.15 dex for abundances, about 10 \% for HI
data and 50 \% for H$_2$. The fact that we used a constant CO-to-H$_2$
conversion factor when we had no abundance gradient for 7 galaxies
must introduce an error in their molecular gas column
densities. However, Fig. 1 of
\citet{boissier03} shows that usually the difference between a
constant and a metallicity-dependent conversion factor is lower than
0.5 dex.
When computing the total gas density (HI+H$_2$), a significant number of points
have only HI data, and the contribution of H$_2$ was then assumed to be 
negligible. This of course introduces another uncertainty on the 
total gas density, however:
i) Most of those points
correspond to outer regions of disks in which CO is detected and
measured only for the inner portion of that same galaxy. It is reasonable
to assume that the molecular fraction is  small in these outer 
regions.  
ii) In 13 galaxies, we only have HI data (even in the center
of the galaxy) but taking into account that only a minor fraction of
the total gas is usually in the form of molecules \citep[about 15 \%,
see ][]{boselli02xco} and that when we do have both profile, H$_2$
dominates only for a few central points, and in most cases by only a
factor a few; we must conclude that the error we commit is relatively small
in most cases. We will nevertheless keep in mind these drawbacks when discussing our results.
\label{secprobh2}

When we found several sets of data, we usually adopted 
one (that seemed better -more recent, better sensitivity),
and checked the consistency between them. In a few cases, we
adopted an average or a combination of several references.
Note that we interpolated the profiles given in the literature
at the resolution adopted in our work. The original resolutions
are very variable, but usually better than the IRAS one.
The gaseous profiles obtained following this procedure  
are given in
the figures in the appendix for each individual galaxy
(at the same resolution as for the rest of our work).
Table \ref{tabrefs} gives the references for all of these data.

\section{Attenuation in the disk of late type galaxies}
\label{secext}
\subsection{Attenuation radial profiles}

The attenuation $A(FUV)$ profiles are shown for each individual
galaxy in the appendix. Most of them present a global decrease
of extinction with radius. 
At large radii, error-bars are very large due to low levels of measured
infra-red emission, so that the data are always consistent with no 
attenuation in the outer disks.
The central value of the estimated FUV attenuation is typically of a
few magnitudes (ranging from virtually 0 to about 6 magnitudes).
The most central point of our profiles could be affected by the
presence of a bulge with e.g. UV upturn, or of a AGN. The fact that we
see in the individual $A(FUV)$ profiles continuous trends with radius
let us think that on average our extinction gradients are not much 
affected by such cases (in the figure of the appendix, only M31 and 
NGC0253 do present a strong disturbance in the central part of the 
$A(FUV)$ profile).

A similar trend of decreasing extinction gradients was present in the
six galaxies of the FOCA/IRAS study of
\citet{boissier04}. \citet{holwerdaV} used counts of distant galaxies,
with a ``synthetic field method'' \citep{holwerdaIII} to estimate
extinction gradients. Their totally independent results are consistent
with ours, although the uncertainties are larger in their method.

At this point, we would like to include some warning concerning the
derived extinction $A(FUV)$. By adopting the \citet{buat05}
calibration, we assume that the star formation histories they adopted
are representative of real galaxies. However, it is probably not the
case for a few galaxies (or at least, parts of galaxies) suffering
heavy quenching of the star formation rate (that could be induced, for
instance by ram pressure in clusters, e.g. Boselli et al. 2006). 
In such galaxies, the GALEX
FUV-NUV color would be intrinsically red (close to or above 1), with
a different relationship between $F_{TIR}/F_{FUV} $ and $A(FUV)$ than
the one we adopted. Such situations will be studied in Cortese et
al. (in preparation). Part of the scatter in the right part of Fig. 1
could be due to such effects. As a result, applying extinction
correction to individual galaxies (or profiles) should still be done
with caution, especially for red FUV-NUV colors. In our approach, the
number of galaxies is sufficiently large so that such effects should
be minor on our global results.

\subsection{The Infrared Excess - UV Slope (IRX-$\beta$) Relationship}

\citet{heckman95} and \citet{meurer95,meurer99}
demonstrated with IUE data the
existence of a relation between the slope of the UV spectrum ($\beta$)
and the TIR / UV ratio (so called IRX-$\beta$ relationship) for
starburst galaxies.
It has been common thereafter (especially for high redshift objects)
to estimate the extinction from the UV slope using the IRX-$\beta$
relationship \citep[e.g.][]{schimi05}.
In the recent years, however, several studies have shown that this
relation does not hold for all galaxies. For example, \citet{bell02}
showed that it fails in nearby galaxies, and \citet{buat05} came to the same
conclusion for samples of UV and far-infrared-selected galaxies.
\citet{kong04} proposed models with a large variety of star formation
histories, demonstrating that they could span a large portion of the
IRX-$\beta$ plane. In this framework, the starburst relation would
hold for galaxies with extremely large current-to-past-average
star-formation rates (birthrate $b$ parameter), while galaxies with less
activity would lie systematically below the relation.
However, it seems that a relation can be  found in nearby galaxies 
\citep[see e.g.][]{gil06,cortese06,seibert05}, given  a simple offset
with respect to the original relation proposed for starbursts.  This
means that the range of star formation histories among real galaxies
is probably not as wide as among models of \citet{kong04} and that the
$b$ parameter might not be the driver of the IRX-$\beta$ relationship.
Actually, \citet{burgarella05} showed that the slope of the extinction and the presence or
not of a bump has a larger impact on the position in the IRX-$\beta$ plane than
the star formation history (and the $b$ parameter).
As discussed in \citet{gil06}, \citet{cortese06} and \citet{seibert05}, the difference
in the IRX-$\beta$ relation for starbursts and normal galaxies might be due to
geometrical effects, aperture effects present in the IUE data.
The relative calibrations in the computation of $\beta$ with IUE's data
and GALEX FUV-NUV color index could also play a role. More work remains to be done.

\begin{figure*}
\includegraphics[angle=0,scale=0.4]{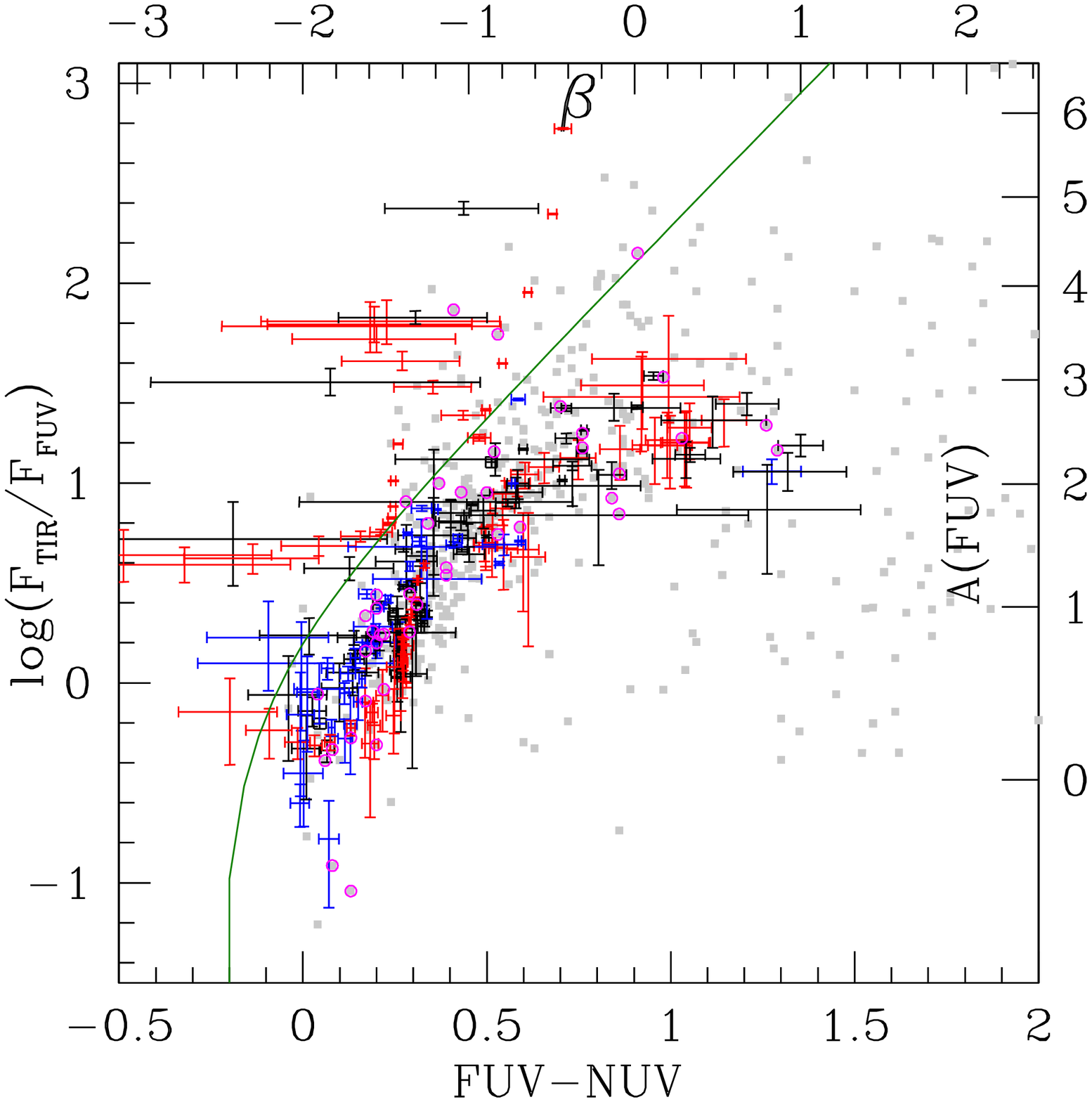}
\includegraphics[angle=0,scale=0.4]{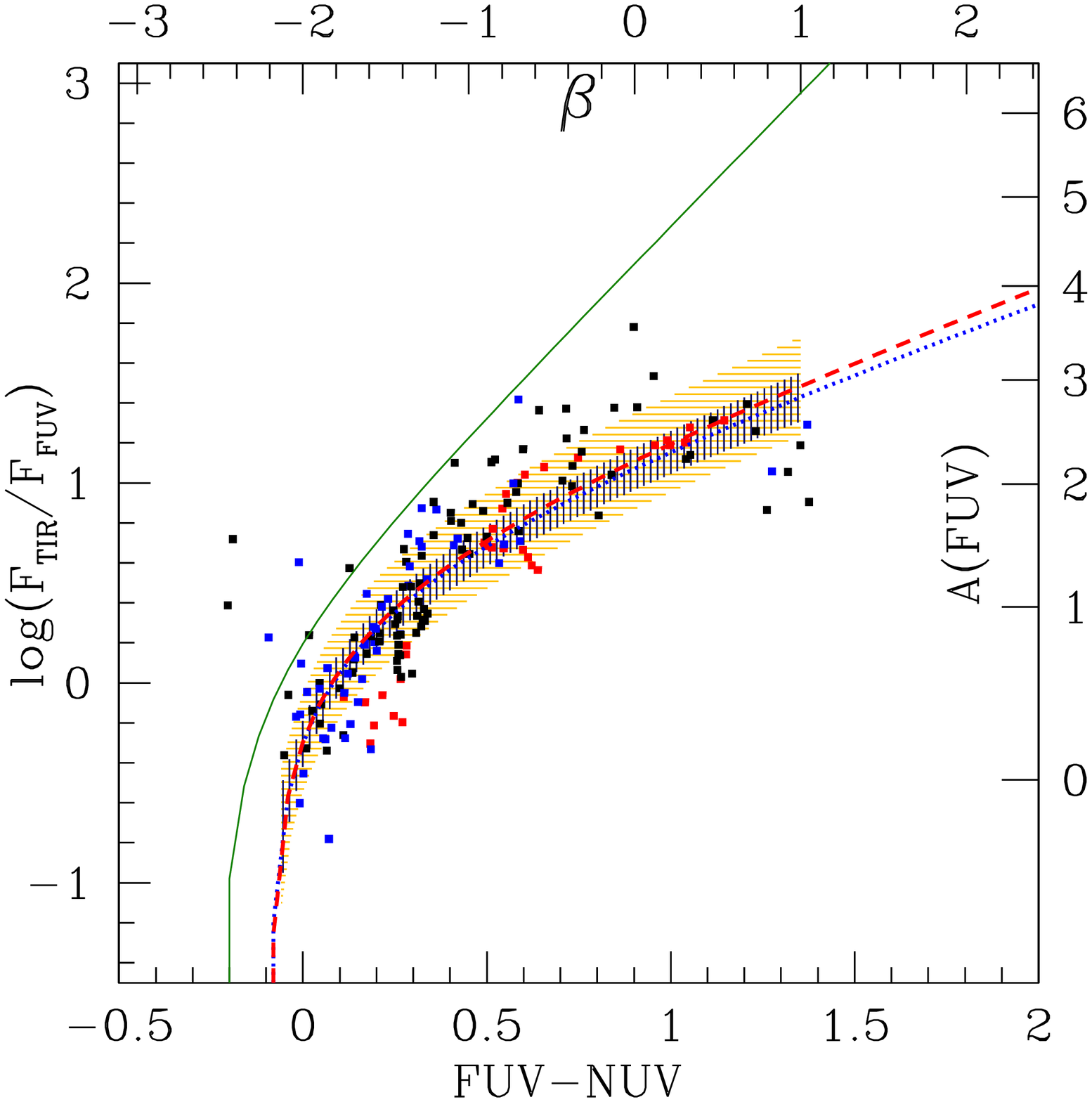}
\caption{\label{figirxbeta} Total infrared to FUV ratio vs 
FUV-NUV color. The FUV-NUV color is a measure of the slope in the UV
$\beta$ given at the top \citep{kong04}.  
The right axis indicates the scale in $A(FUV)$ using equation \ref{eqbuatcal}.
In the left panel, points with errorbars larger than 0.5 
(dex on left axis and mag on bottom axis) were omitted for the sake of 
clarity (errorbars indicate how much the results would change
if our sky determination in the UV or far infrared was
moved by $\pm 1 \sigma$).
The line is the usual relation for startbusts as given in
\citet{kong04}. (Electronic version: different colors indicates
various axis ratio b/a: red for $0<b/a<1/3$; black for $1/3<b/a<2/3$;
blue for $2/3<b/a<1$).  Grey squares are the integrated values of the
GALEX Atlas of nearby galaxies \citep{gil06}. Squares with a circle
correspond to the galaxies in our study.
In the right panel, we show as black squares the points used to
compute our fits (excluding nearly edge-on galaxies). The usual
starbursts relationship is compared to our fit and its 1 sigma
deviation for a simple fit (dashed curve and horizontally hatched area)
and a fit taking into account the uncertainties (dotted line and
vertically hatched area).}
\end{figure*}

In this context, our nearby galaxies have the advantage of being close
enough so that we can spatially resolve the IRX-$\beta$
relationship. While this has been done by looking at individual
regions in a few galaxies already, for instance M51
\citep{calzetti05}, we present for the first time the IRX-$\beta$
relationship for the profiles of all the large late-type galaxies of
the GALEX Nearby Galaxies Atlas in Figure \ref{figirxbeta}.
It is particularly interesting to note that while outliers are present
(with correspondingly larger error-bars), most of the points describe
a very tight relation, but shifted from the classical relation derived
for/from starbursts, similar to the recent works of
\citet{gil06,cortese06,seibert05} (see discussion above for the
possible reasons for this shift).
In order to quantify the relationship, we performed a
nonlinear least squares fit of the form
$y=log(10^{a+b x}-c)$ (with $x$=FUV-NUV, the color between the two galex
bands, and $y$=$log(F_{TIR}/F_{FUV})$, after
rejecting quite edge-on galaxies(NGC~0055, NGC~0253, NGC~0660,
NGC~0891, NGC~4216, NGC~4621, NGC~4656a). We obtained for $a,b,c$
respectively 0.570, 0.671 and 3.220 using y-uncertainties as weight,
and 0.561, 0.713, 3.136 using no weights. These fits and their associated
1 sigma dispersion are shown in the right part of Fig. \ref{figirxbeta}.

The fact that our IRX-$\beta$ relationship is quite tight is also
interesting as there is much more scatter when individual regions are
looked at, see for instance \citet{calzetti05} in M51a, or
\citet{thilker06} in NGC~7331, both studies based on the GALEX/SINGS 
joint analysis of GALEX and Spitzer data from the SINGS project
\citep{kenni03b}.
By azimuthally averaging profiles, we remove the
effects due to small-scale star-formation histories (affecting the UV
slope), and is likely to reduce the effect of radiative transfer
peculiarities on small scales: that is, the dust in some small regions
may well be heated by the UV emission from neighboring regions.  Using
the low resolution of IRAS (coupled with azimuthally averaging) may
turn out to be an advantage in this respect.

Figure \ref{figirxbeta} also presents the integrated values for the galaxies
of the Atlas \citep{gil06} as grey squares. The squares overploted
with a circle correspond to the integrated value given in the Atlas
for the galaxies of the present study.
The relation of our study seems tighter than the one for the
integrated galaxies of the Atlas.
This can be explained by the combination of several effects:
i) For all our galaxies, we followed exactly the same
procedure from scratch, i.e. we do not use published fluxes
but rather re-derive them ab initio, in contrast the UV Atlas for which
fluxes were compiled from a wide variety of sources.  ii) In
integrated galaxies, centers with activity 
might affect the whole flux if they are strong enough, while they
probably influence only the central point in a profile like the one we
compile.
About half of our
galaxies have a sign of central activity (AGN, LINER, Seyfert
indicated in NED), but precisely because it is confined to the central
point, it should not affect much our profiles.

\subsection{The Absence of a trend between Attenuation and Gas Column Density}

\begin{figure*}
\includegraphics[angle=0,scale=.40]{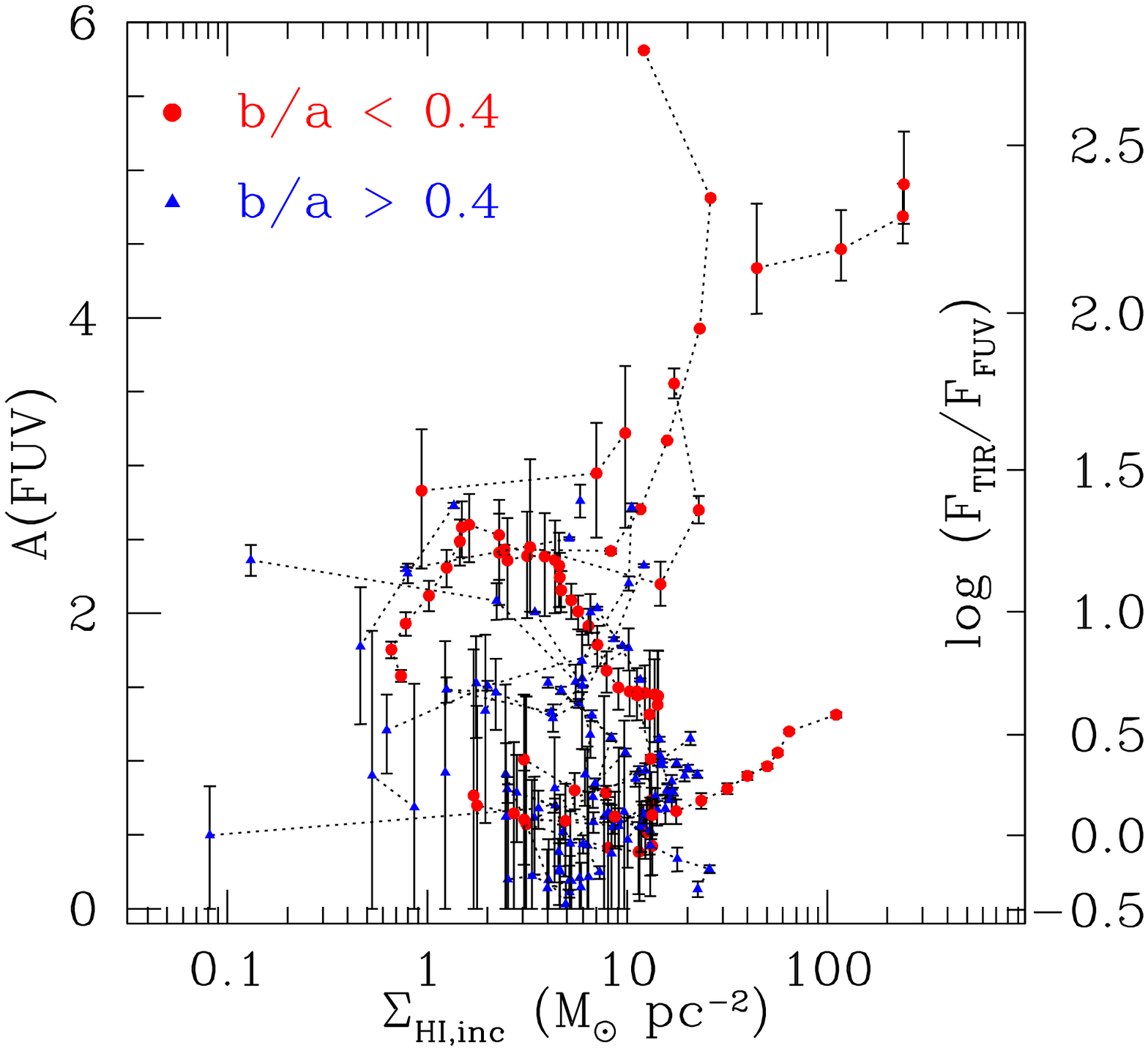}
\includegraphics[angle=0,scale=.40]{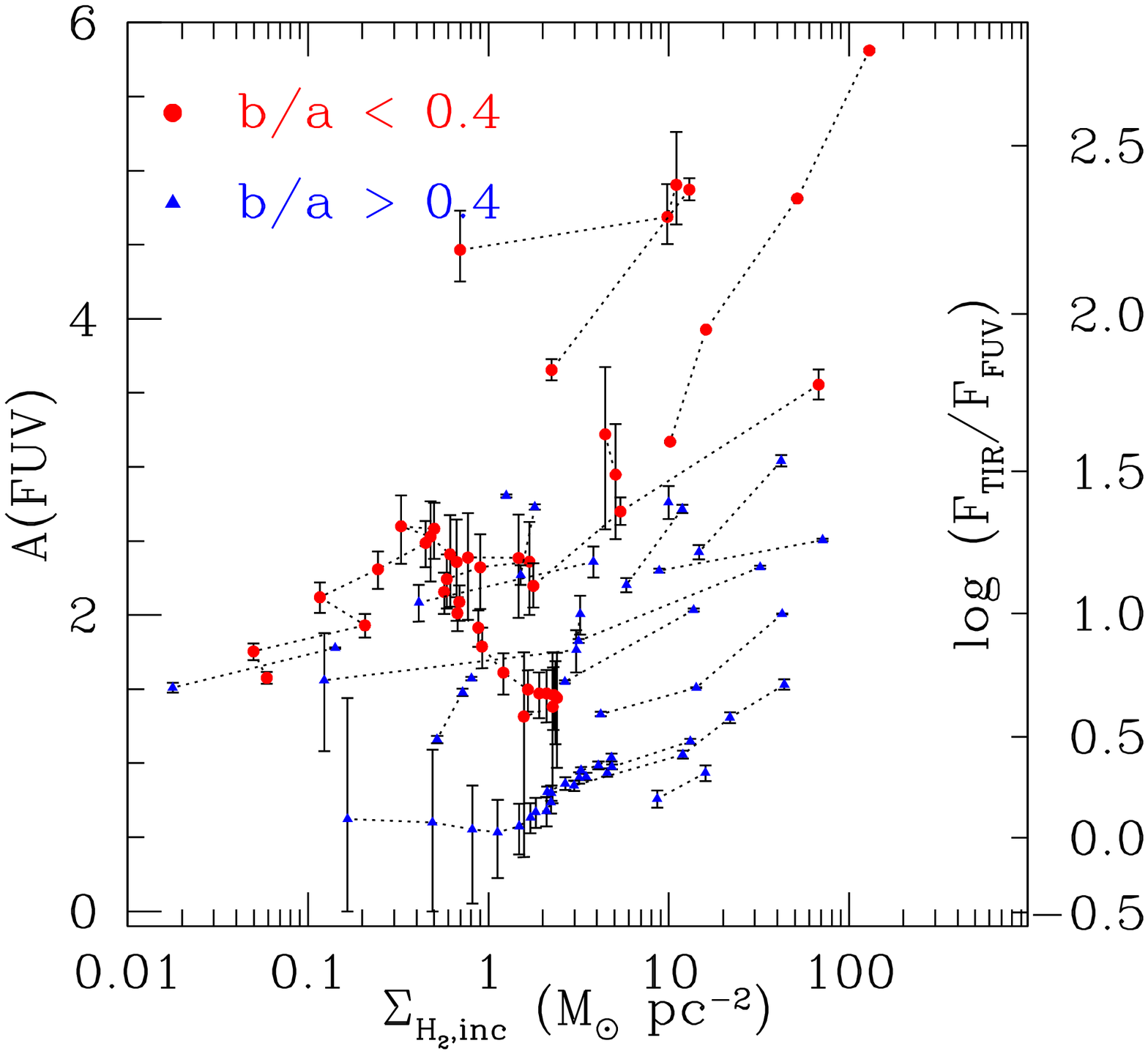}
\caption{\label{figAn}Attenuation at FUV wavelength 
($F_{TIR}/F_{FUV}$ ratio scale on the right axis) as a function of the
observed column density of HI (left), H$_2$ (right). Points belonging
to a same galaxy are connected by a dotted line. Different symbols
indicate different values of the axis ratio (b/a) as indicated in the
corner.  Note that since $A(FUV)$ is not corrected for inclination,
the gas column densities used correspond to the same inclination
(hence the ``inc'' subscript).
Errorbars indicate how much the results would change
if our sky determination in the UV or far infrared was
moved by $\pm 1 \sigma$.}
\end{figure*}

\begin{figure}
\includegraphics[angle=0,scale=0.4]{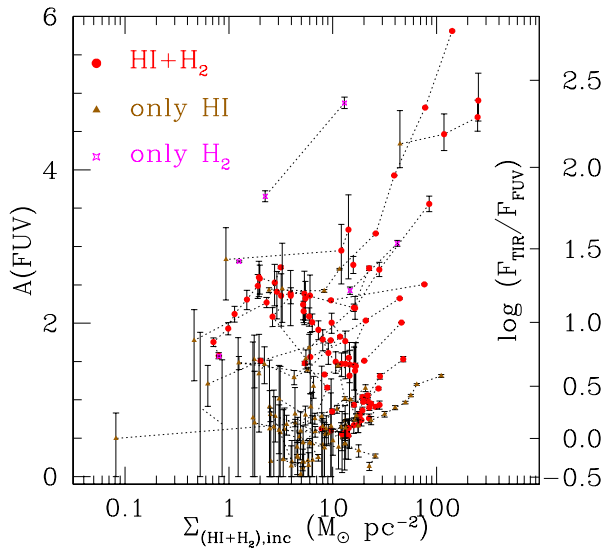}
\caption{\label{figAtot}Attenuation at FUV wavelength 
($F_{TIR}/F_{FUV}$ ratio scale on the right axis) as a function of the observed
total gas density (HI+H$_2$).  Since $A(FUV)$ is not corrected for
inclination, the gas column densities used correspond to the same
inclination (hence the ``inc'' subscript). Errorbars indicate how much the results would change
if our sky determination in the UV or far infrared was
moved by $\pm 1 \sigma$.  }
\end{figure}

It is relatively common to consider the attenuation to gas density
ratio as being constant. For instance,
\citet{komugi05} adopt $N_H = 2 \times 10^{21} A_V$. This is
essentially inspired by studies in the Milky Way such as
\citet{bohlin78}.  Under this assumption, one would expect a strong
correlation between gas density and attenuation, what is not 
supported by our results (see Fig. \ref{figAn}, \ref{figAtot}).
Thus, attenuation (and therefore the existence
of dust) is probably affected by various processes not directly related
to the gas density.  In Fig. \ref{figAn}, we show the amount of
extinction as a function of the gas in the form of neutral hydrogen
(HI, left) and in the form of molecular hydrogen (H$_2$, right).
Within individual galaxies (points connected by one dotted curve), a trend
of increasing attenuation with gas densities is generally visible (specially 
with H$_2$). However, the differences from one galaxy to another are large, 
so that globally no relation is clearly emerging.
The fact that the amount of attenuation does not seem to correlate
with the HI colum density was also found by \citet{holwerdaV}.
A small trend can be seen between the molecular gas and the
extinction. While this seems reasonable (inner regions are denser, have
larger extinction and higher molecular fraction) and is observed in
most galaxies individually, the variations from one galaxy to another
are still extremely large. 
In Fig. \ref{figAtot}, we compare the ultraviolet attenuation $A(FUV)$
to the total gas density. In this case, we are in an intermediate
situation.  At low gas column densities, which correspond usually to
the outer part of galaxies (where errors are larger), no trend is
observed. In the inner parts of galaxies, the situation is the same as
for H$_2$ since the molecular gas often dominates the total gas
density or at least appears to compensate for the central plateau or
the observed decrease of HI in the central regions of galaxies.

In conclusion, our data do not support any relation of the type $A/N_H
= constant$ (commonly used in the literature). This is actually not
surprising as $A(FUV)$ is not a measure of the dust mass, but of the
amount of dust heated by nearby young massive stars. As such, it will
be largely affected by geometrical effects, and the physical
properties of the dust grains (type, size distribution...), what would
probably erase any underlying relation between the dust and gas
masses. Because the dust mass is dominated by cold dust, $A(FUV)$ is
just not a good measure of it. Estimating the dust mass, and thus the
dust to gas {\it mass} ratio is out of the scope of this paper since IRAS
wavelengths do not allow us to probe the cold dust.

\subsection{The Attenuation-Metallicity Relationship}

\label{secAz}

\begin{figure}
\includegraphics[angle=0,scale=0.4]{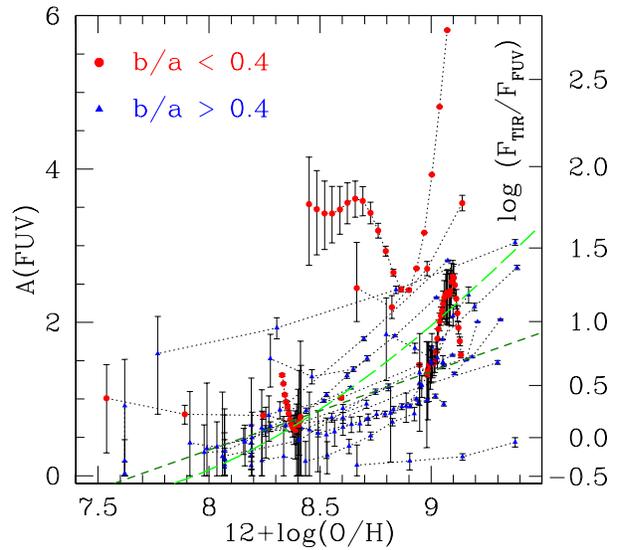}
\caption{\label{figAz}Attenuation at FUV wavelength
($F_{TIR}/F_{FUV}$ ratio scale on the right axis) as a function of the
metallicity (as given by the abundance gradient of each galaxy).
The short-dashed line is a fit (A(FUV)=1.02 (12+log(O/H)) - 7.84) to
the points belonging to galaxies with b/a $>$ 0.4 (see section \ref{secfitextz}). 
The long-dashed
curve corresponds to the fit of $F_{TIR}/F_{FUV}$ vs O/H in
\citet{cortese06}. Errorbars indicate how much the results would change
if our sky determination in the UV or far infrared was
moved by $\pm 1 \sigma$. }
\end{figure}

While the relation observed with the gas is ill-defined, the
attenuation seems to depend in a clearer way on the metallicity
(Fig. \ref{figAz}).  Indeed, for a given abundance of oxygen, the
scatter in observed extinction is smaller than at any given gas
surface density. This is especially true if we exclude the galaxies
with the largest inclinations in which the method used to derive the
attenuation might suffer from various problems as commented in section
\ref{secprobincli}.  We performed a least squares minimization 
fit to the points within galaxies
with axis ratio larger than 0.4 and found:
\label{secfitextz}
\begin{equation}
A(FUV)=1.02 \times (12+log(O/H)) - 7.84.
\end{equation}
(the dispersion around this fit is $\sigma_{A(FUV)}$=0.5 mag, 
and the correlation coefficient is 0.65).

\par\noindent This is qualitatively consistent with the earlier work
of \citet{boissier04}.  Albeit a bit shallower, this relation is
characterized by similar attenuations in the high  metallicity range
(at $12+log(O/H) \sim$ 9.3). Due to the limited number of galaxies,
there were no points at the lower metallicities (7.5-8.5) in
\citet{boissier04}. A source of potential difference between this
earlier study and the present one is that it was performed at the FOCA wavelength
rather than GALEX.
In \citet{boissier04}, the extinction determined in the integrated
star forming galaxies of \citet{buat02} with metallicities of
\citet{gavazzi04} were also shown to exhibit a similar trend.
Our results are also in good agreement with the recent study of
\citet{cortese06} (although their trend is slightly stronger than
ours), for a sample of integrated cluster galaxies observed with GALEX (their
fit of the TIR-to-FUV ratio is converted to our $A(FUV)$ scale in
Fig. \ref{figAz}).

Note that the attenuation-metallicity relationship we found concerns
galaxies forming stars in a relatively quiescent (disk)-mode.
\citet{heckman98} have shown that the UV slope is correlated with
the metallicity in starbursts, also corresponding to a metallicity-extinction
relationship among starburst, shifted however with respect to the 
spirals to higher extinctions  \citep[and presented in ][]{boissier04}.

\section{Star Formation Law}
\label{secstarform}

\subsection{The Theoretical Star Formation Laws}

The star formation on galactic scales is obviously a crucial phenomenon 
in the evolution of galaxies. As a result, it is also a fundamental element
of any galaxy evolutionary model, e.g. N-body/semi-analytical models like
GALICS \citep{hatton03}, SPH/chemical evolution models \citep{lia02},
chemical / spectrophotometric evolution models  \citep{boissier99}.
Despite its importance, only rough theories exist and a few 
empirical relationships have been looked for 
\citep[e.g.][]{madore77,buat89b,kenni98a,wongblitz}.
An excellent review of the empirical situation is given in
\citet{kenni98b}, while \citet{elmegreen02} consider the possible
physical origin of the empirical laws.
The reader is refereed to these works for an extensive discussion on the topic.
Here, we only come back rather quickly on a few of the relations proposed
for the ``star formation law''.
These ``laws'' should allow us to  
predict the star formation rate from other physical quantities,
as shown below.
Especially, we will look for relations between the star formation rate
surface density ($\Sigma_{SFR}$) and the gas surface density in its
various phases (neutral, molecular) as well as in total gas
(neutral + molecular), under the form of a traditional ``Schmidt law''
(see Schmidt, 1959 for the original work):
\begin{equation}
\Sigma_{SFR} = \alpha  \Sigma_{GAS}^{n}. \label{eqschmidt}
\end{equation}

It is reasonable to expect a more direct relationship with the
molecular gas rather than the neutral gas since this phase is more
closely in the sequence related to the ensuing star formation
event. For instance, in their disk model consisting of
self-gravitating clouds interacting to produce a turbulent viscosity,
\citet{vollmer02} found that the star formation rate and molecular gas
surface densities should follow the same dependence with galactic
radius. On the observational side, while a relation between star
formation rate and molecular mass was not initially 
confirmed \citep{kennicutt89,boselli95},
\citet{boselli02xco} showed that using a luminosity or
metallicity-dependent CO-to-H$_2$ conversion factor provides a
better relation than using HI alone.

While this suggests a more direct connection between the star formation
rate and the molecular gas, the cause for star formation could still
be related to the total gas.  Indeed, a basic model of self
gravitating disk would suggest a Schmidt law with the total gas
density as the main control parameter \citep{kenni98a}.

The other model we will consider here is a Schmidt law modulated by a
dynamical factor:
\begin{equation}
\label{eqsfr}
\Sigma_{SFR} = \alpha  \Sigma_{GAS}^{n} \frac{V(R)}{R}
\end{equation}
where $V(R)$ is the rotational velocity at radius $R$ (where the
surface densities are determined). The ratio $V(R)/R$ can be
considered either as representing the frequency of the passage of
density waves enhancing star formation \citep{Ohnishi75,wyse89} or a
dynamical time-scale \citep{kenni98b}. In their model of the chemical
evolution of the Milky Way, \citet{prantzos95} found that an another
factor in addition to a simple Schmidt law was necessary and opted for
a relation of the type given by equation \ref{eqsfr} with $V(R) = $ a
constant, and $n =$ 1.
Subsequent models of the Milky Way and spirals used a similar 
SFR law \citep{boissier99,boissier00} but with $n=1.5$. This 
was found to give  agreement with
the H$\alpha$ profiles of 16 nearby spirals studied by
\citet{boissier03} although other laws could not be
excluded.
\citet{boissier03}, had also tested a dependence on the stellar
surface density in addition to the gas, as proposed by
\citet{dopita94}. Despite having an additional free parameter, this
formulation did not offer a better fit to their data. To explore this
further we would need near-infrared surface brightnesses for our
galaxies. Unfortunately 2MASS data are not sufficiently deep in outer
parts of our galaxies, and we decided to omit consideration of this
law in our present study.

Before testing the above proposed laws, we should note that the field
of galactic star formation is not lacking new ideas.  For instance,
\citet{elmegreen05} proposed that a combination of processes trigger
star formation, conforming to the observed relations and thresholds
(densities below which no star formation would occur).
\citet{seigar05} presented an interesting correlation between the
specific star formation rate and the shear rate, stressing the
possible role of shear in star formation; however the shear versus
average star formation rate (his Figure 4) has large dispersion. Our
rotation curves do not have the necessary resolution or homogeneity to
explore trends involving the shear rate.
It has also been recently suggested \citep{barnes04} that star
formation may depend not only on the local gas density but may also be
induced by shocks, especially in the case of interacting galaxies
(we do not have such violently interacting systems in our sample).
Testing this kind of star formation law requires detailed dynamical
modelling on a case by case basis.  \citet{barnes04} showed that this
idea is promising by reproducing observations of the Mice. It would be
however hard to implement such a law in e.g. 
semi-analytic model since it needs a detailed dynamical modelling of
each interacting systems (including their precise geometry and
details of the encounter).

In any case, a best possible law for star formation should still properly
reproduce the relation observed between gas, dynamics and star
formation indicators that we compiled for our GALEX galaxies, whatever
the physical causes are. Thus, this purely empirical work provides
additional constraints for future theoretical work and galaxy
evolution models.

\subsection{Our constraints on the star formation laws}

\begin{figure}
\plottwo{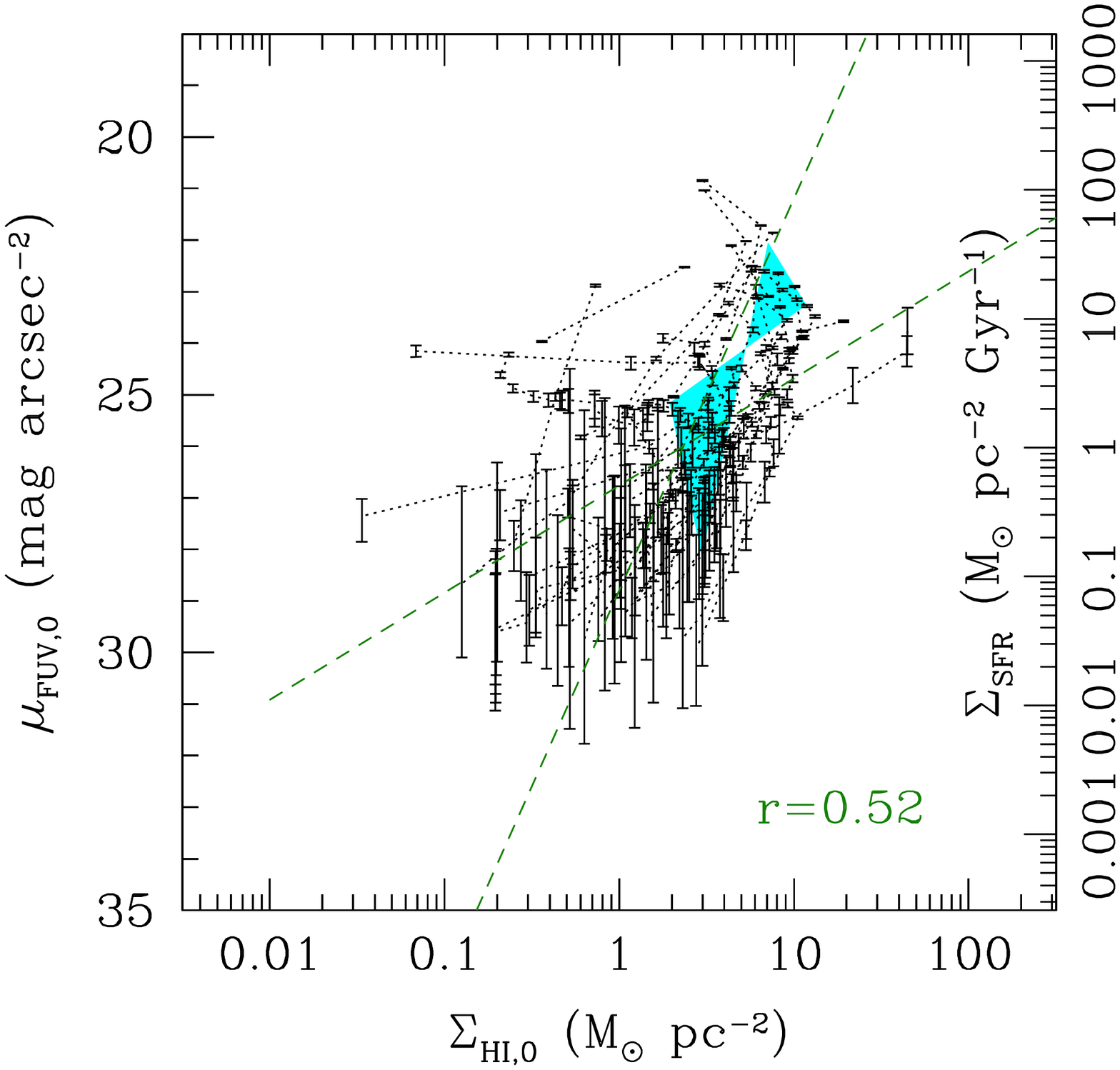}{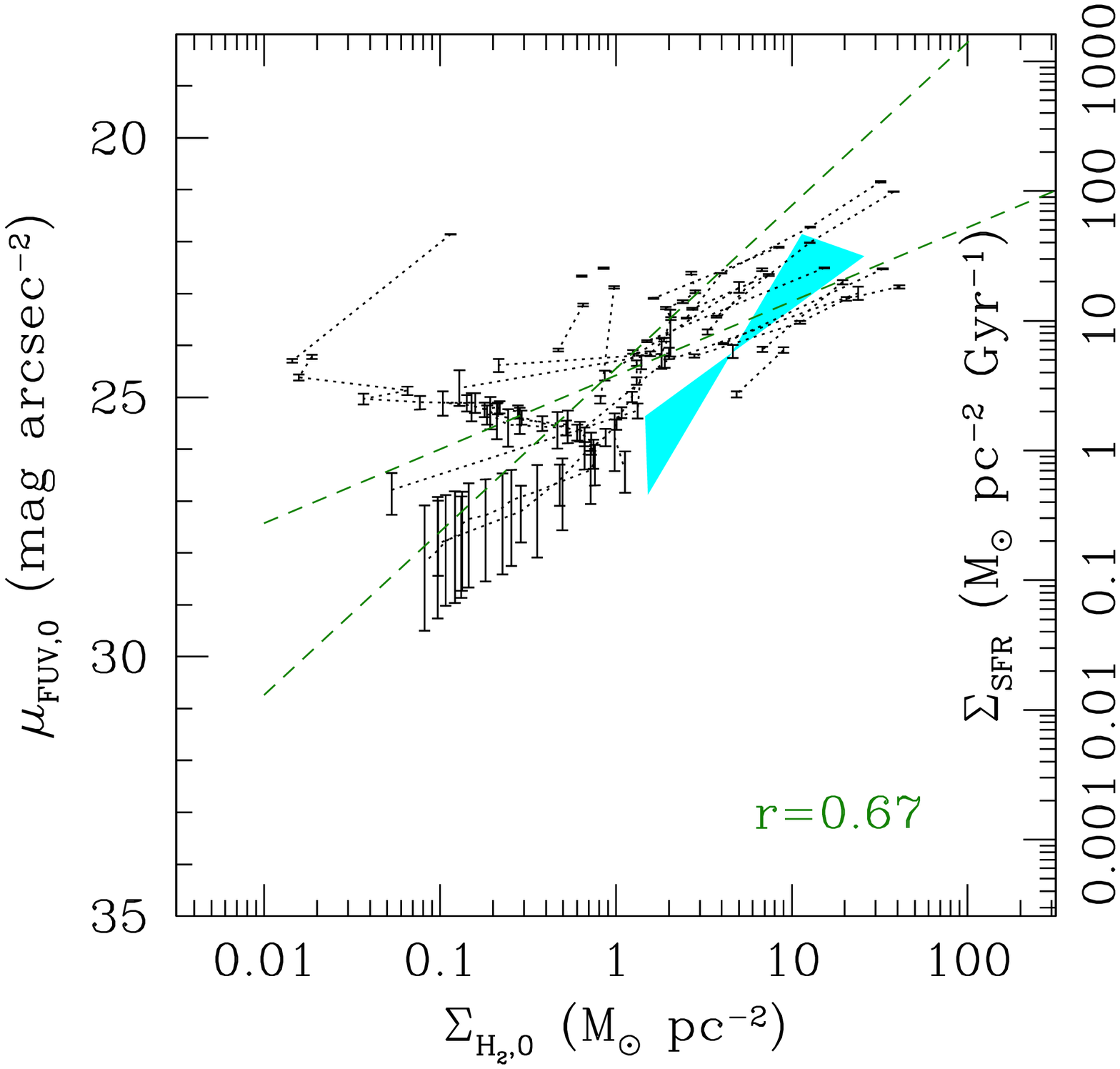}

\plottwo{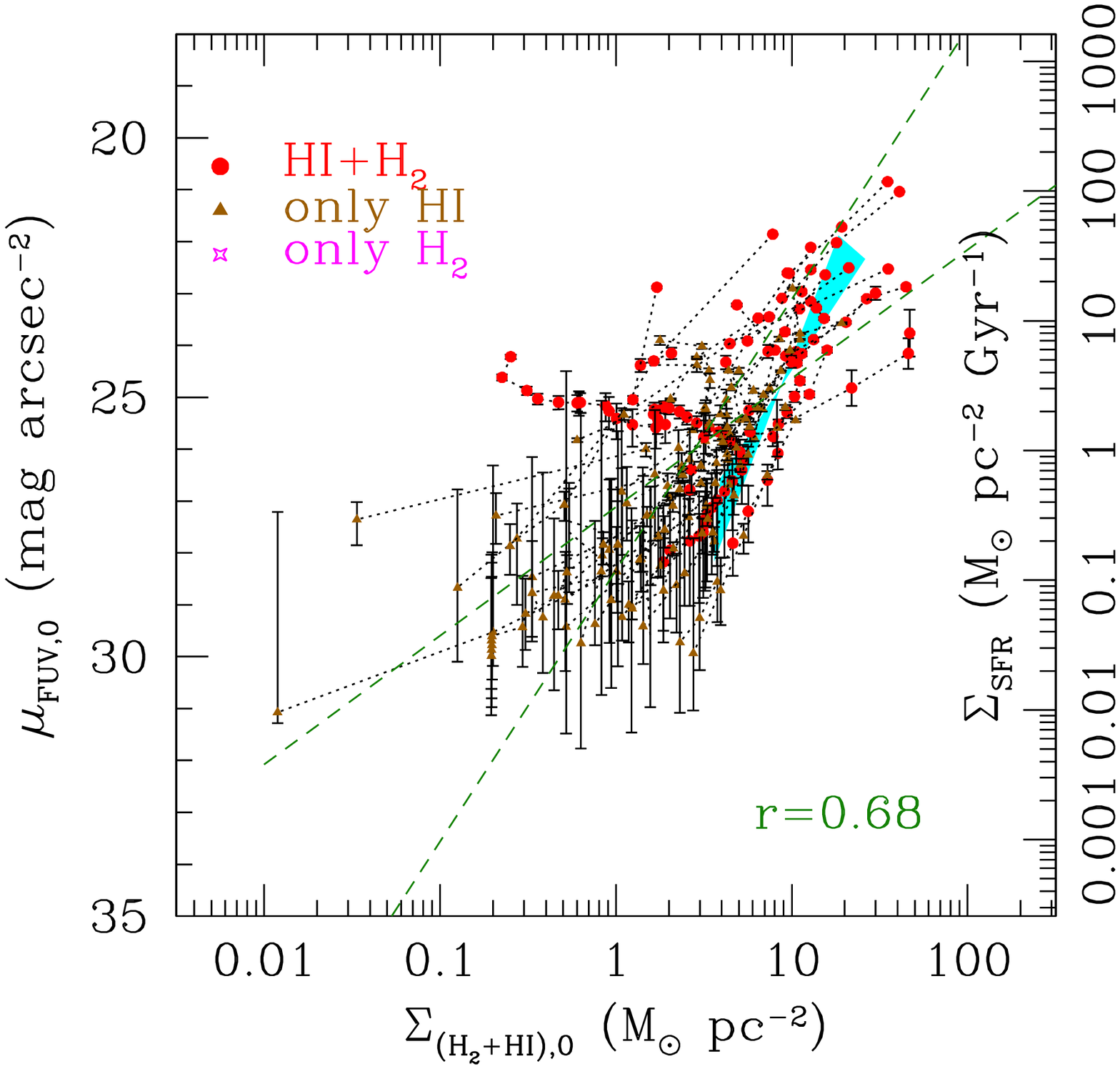}{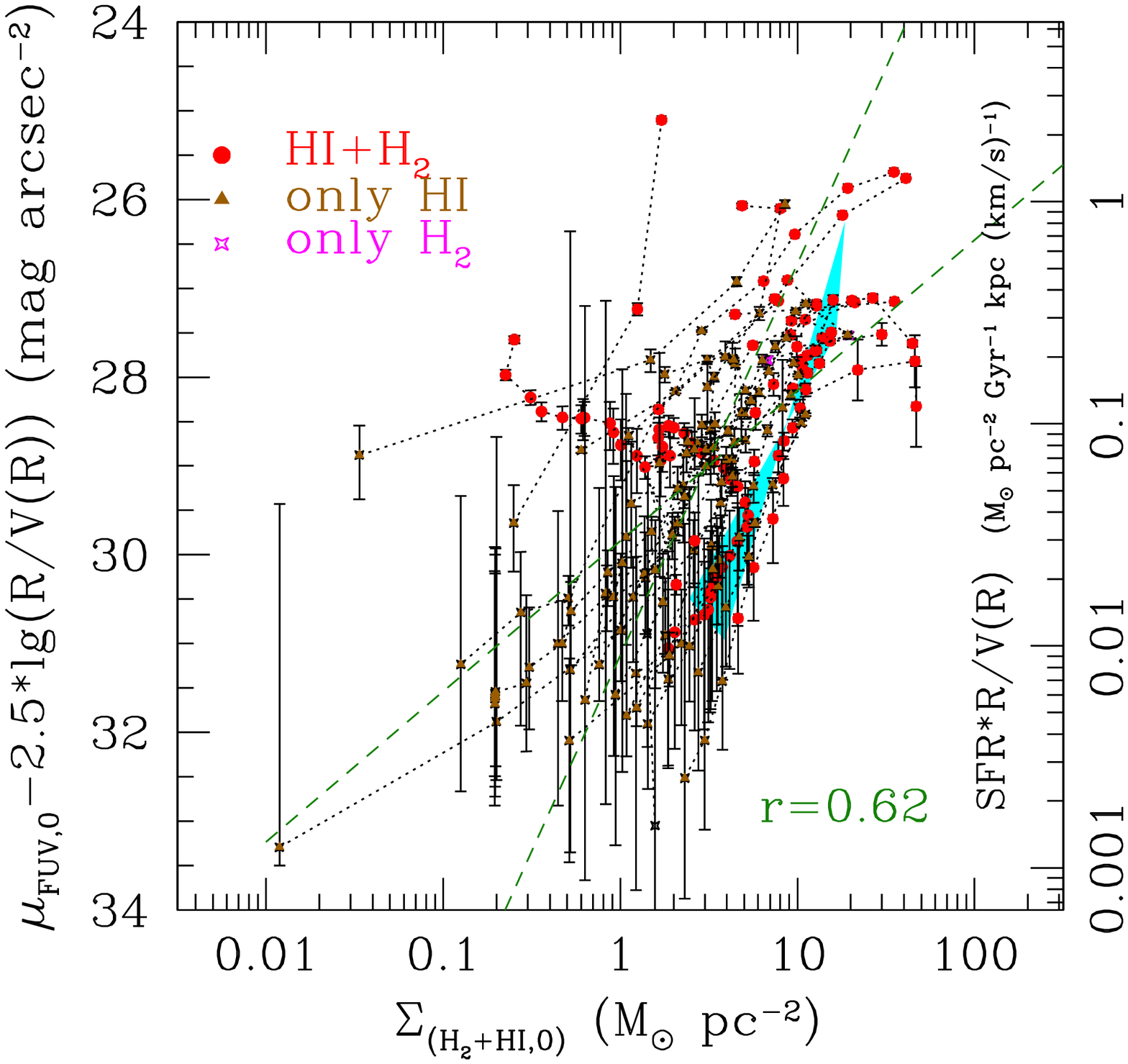}

\caption{\label{figsfrlaws} Test of various ``Schmidt laws'' for 
our whole sample. The star formation rate (right axis) derived from FUV
profiles corrected for attenuation and inclination is compared to 
HI (top left), H$_2$ (top right), total H surface densities (bottom-left). In the bottom-right
panel, we show the SFR $\times$ R/V(R) vs hydrogen total density to test equation
\ref{eqsfr}. In each box, the hourglass shaped shaded area indicates equivalent
results from \citet{boissier03} based on H$\alpha$ profiles (the
extent corresponds roughly to the extent of the observations and the
diagonals to the regression lines of this study).
The correlation coefficient and regression lines (dashed lines,
obtained fitting y vs x and x vs y, see section \ref{secfitsfrlaw}) 
are shown (their parameters are given in table \ref{tabregline}).
Errorbars indicate how much the results would change
if our sky determination in the UV or far infrared was
moved by $\pm 1 \sigma$.
}
\end{figure}

In each panel of Fig. \ref{figsfrlaws} we show a variant of the
simple Schmidt law (SFR versus neutral gas, molecular gas, total gas),
testing equation \ref{eqschmidt} and
a combination of the SFR + dynamical factor versus total gas
testing the relation of equation \ref{eqsfr}.  In this figure, the
internal contributions of individual galaxies are connected by dotted
lines. Individual SFR and gas profiles, as well as SFR vs gas are
given for each galaxy individually in the Appendix. Correlation
coefficients are indicated in each panel, as well as the two (dashed)
regression lines whose coefficients are given in Table
\ref{tabregline}. For each relation, the first regression line is obtained
by performing a least square fit of the star formation rate surface densities
with the gas surface densities being given (minizing the errors on the Y axis), 
and the other one by performing a least square fit of the gas surface densities
with the star formation rate surface densities being given (minimizing the
errors on the X axis). \label{secfitsfrlaw}
\clearpage
\begin{deluxetable}{r r r r r r r r }
\tablecaption{\label{tabregline}Parameters $a_i$ and $b_i$ (i=1,2) 
of the two regression lines ($log(Y)=a_i$ $log(X)+b_i$) shown as 
dashed lines in each panels of Figure \ref{figsfrlaws} (least square fit minimizing the 
errors on X and Y respectively). We also give for each set of variables ($X$,$Y$)
the correlation coefficient $r$ and the number of points used for the fit ($N$).}
\startdata
$X$               &    $Y$                            & $a_1$ &   $b_1$ & $a_2$ & $b_2$ & $r$   & N   \\ 
$\Sigma_{HI}$     & $\Sigma_{SFR}$                    & 0.83  &  -0.30  & 3.04  & -1.11 & 0.52  & 232 \\
$\Sigma_{H_2}$    & $\Sigma_{SFR}$                    & 0.57  &   0.58  & 1.26  &  0.63 & 0.67  & 108 \\
$\Sigma_{HI+H_2}$ & $\Sigma_{SFR}$                    & 0.99  &  -0.44  & 2.09  & -0.92 & 0.68  & 236 \\
$\Sigma_{HI+H_2}$ & $\Sigma_{SFR}\times $R / V(R)$ $  & 0.68  &  -1.53  & 1.76  & -2.04 & 0.62  & 228 \\ 
\enddata
\end{deluxetable}
\clearpage

It is obvious from this figure that the form of the observed star
formation laws differs from one spiral galaxy to another. This was
already known and discussed earlier
\citep[e.g.,][]{boissier03,kenni98a}. It is hoped that averaging over
many galaxies is equivalent to averaging the radial star formation
history of any given galaxy over time.
The trend with H$_2$ seems relatively good on the high density side
(larger than $\sim$ 1 M$_{\odot}$ pc$^{-2}$) but
the number of points with molecular gas alone is very small, and the
results are inconclusive, especially given the additional uncertainty
on the conversion factor. As in the case of extinction vs H$_2$,
however the scatter seems too large to be caused by this factor alone.
We note that \citet{komugi05} show that H$_2$ is very well correlated
with the star formation rate but they are interested in high-density
regions and galactic nuclei, working at high resolution. In our case,
to the contrary, the central parts of the galaxies have little
influence in our plots since we are working at low resolution and
looking at the largest possible radii.
When the total gas is used, as in the study of the attenuation, we
acknowledge that the absence of molecular gas data can affect the
precise value of the slope and the intercept; however, the scatter
seems too large to be entirely due to the uncertainties in the H$_2$
content alone. \citet{boissier03} showed that the extrapolation (or
not) of CO data, and the use of a metallicity-dependent or a constant
conversion factor did change significantly their results, but mostly
influenced the star formation threshold (without much affecting the
rate exponent $n$).

Among all of the Schmidt law variants that we tried, the simple
Schmidt law (with a dependence on the total gas density) gives the
best results, although the correlation coefficient for the molecular
gas and the ``dynamical'' law (respectively 0.67 and 0.62) are close
to the one for the total gas (0.68).
\citet{komugi05} have shown that the simple Schmidt law
extends to higher gas surface densities when smaller and central
regions (assumed to be dominated by H$_2$) are considered. If added 
to our diagrams, their data would go from about 10 to 1000
$\rm M_{\odot} pc^{-2}$ for the gas, corresponding to a range in star
formation rate from about 1 to 1000 $\rm M_{\odot} pc^{-2} Gyr^{-1}$.
This is compatible with an extrapolation of our adopted Schmidt law (using the
total gas) to larger surface densities, with a similar slope. We
notice however a shift to lower star formation rates for the same gas
density.  In their study, the SFR is derived from H$\alpha$, which
might explain this offset (see below for a discussion on this point).

In all panels of Fig. \ref{figsfrlaws}, we also show for
comparison the results of \citet{boissier03}, in the form of a
hourglass-shaped shaded area (the extent of the hourglass being the
extent of the observations, and the diagonals the regression lines
found in this work).  This study used H$\alpha$ profiles for 16 spiral
galaxies.  There are some systematic differences: on average they
found lower star formation rates for the same gas density than in our
present study, except with HI. The slope of a simple Schmidt law was
also slightly steeper than what we find.  Overall, taking into account the
large dispersion, the differences are not dramatic and may be caused
by i) inadequate/uncertain corrections: for instance, in \citet{boissier03}
H$\alpha$ was corrected for extinction and [NII] contamination by
standard factors depending only on the galaxies type (a common
procedure), but not considering possible radial variations, ii) calibration
issues : the conversion of H$\alpha$ and UV light to star formation
rates depends on the initial mass function adopted for converting the
observed fluxes \citep{kenni98b}. For example, if the assumed IMF
overestimates the number of ionizing stars, the star formation rate
derived from H$\alpha$ will be underestimated.
Another difference, that cannot be explained by these effects is the fact that
the gas extent was smaller, i.e. with our UV profile, we detect low levels
of star formation, corresponding to lower levels of gas surface densities.

\begin{figure}
\includegraphics[angle=0,scale=.40]{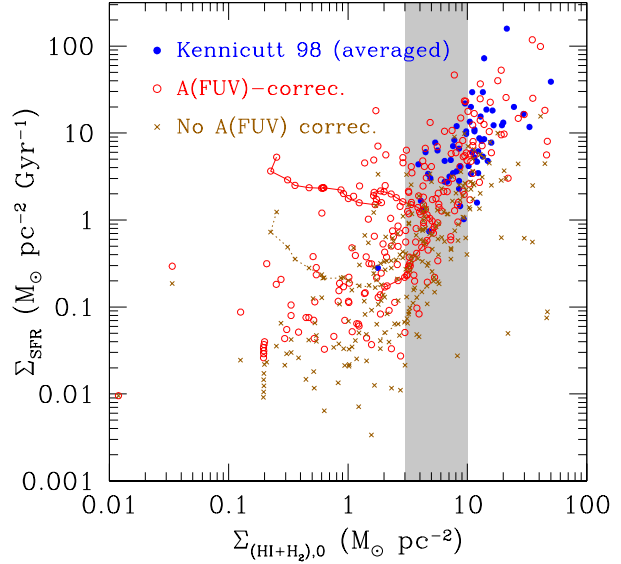}
\caption{\label{figkenni}The simple Schmidt law (Star formation rate surface density vs 
total Hydrogen surface density) found in our work along radial
profiles of GALEX galaxies (including the extinction-correction: open
circles, or not: crosses) is compared to the one of \citet{kenni98a}
(filled circles, obtained by computing average surface densities
within the optical disk). The shaded area indicates gas densities values
found at the threshold by \citet{kennicutt89}.
The studies overlap with each other in the
high density half, but we obtain many more points at low surface
densities. The points corresponding to M31 (following an untypical path
in the inner galaxy) are connected: high SFR and low gas amounts (top-left) 
correspond to the inner galaxy. When increasing the radius, the SFR decreases 
while the gas increases (contrary to classical SFR ``laws''). At about 50 arcmin,
the trend changes and both the SFR and the gas surface densities go down with radius,
following the SFR ``law'' described by other galaxies.}
\end{figure}

In Fig. \ref{figkenni}, we compare the simple Schmidt law (obtained
from our UV radial profiles versus total gas) to the galaxies of 
\citet{kenni98a}, obtained from H$\alpha$ imaging of
spiral galaxies, averaged within their optical disk.  
Considering the scatter, the agreement between both studies is quite
good. Here again, however, we have many more points at low gas
densities. We also include in this diagram the star formation rate
values derived from the UV assuming zero attenuation (crosses). These
are lower limits on the star formation rate, showing that a significant
amount of star formation is detected at low gas densities, and this
result does not depend on the attenuation correction.
This comes from several facts: i) Kennicutt's estimates are
averages within R$_{25}$, where densities are on average higher ii) In
the UV, we do find emission at low levels of star formation (in stark
contrast to H$\alpha$ observations), corresponding to gas surface
densities below the putative threshold for star formation.

In this figure, we also connected the points corresponding to M31. 
The inner 50'' of this galaxy are orthogonal to the 
usual Schmidt relation (and the one found in the rest of the sample).
We leave for later studies to interpret this untypical behavior
of one of our nearest neighbor.

Our plots might appear more highly dispersed than other studies
\citep[e.g.][]{komugi05,kenni98a} but this is mainly because we do not
work on a large range of gas densities (i.e., the scatter only seems
smaller in \citet{kenni98a} because of the larger dynamic range
introduced into the plots when circumnuclear starbursts are included),
and we are working primarily at low densities where stochastic effects
and larger uncertainties may have a larger impact.  Another possible
cause of the dispersion is that we are using UV data, and as
\citet{buat89b} noted, the time-scales we are probing is significantly
longer than with H$\alpha$, probably making the underlying ``causal''
relations harder to catch \citep{madore77}.

\subsection{Concerning the Notion of a Threshold}

Star formation rate profiles, derived from H$\alpha$ data
are known to frequently present an abrupt break, while 
atomic gas is generally much more spread than the star forming and 
the stellar disc
\citep[e.g.][]{martin01}. This observation is traditionally 
interpreted as the result of local instabilities developing in 
differentially rotating discs within a threshold radius
where the gas density is larger than a critical density
\citep{toomre64}. More complex physics has also been suggested
\citep[e.g. the onset of thermal instability of][]{joop04}.

\begin{figure*}
\includegraphics[angle=0,scale=.80]{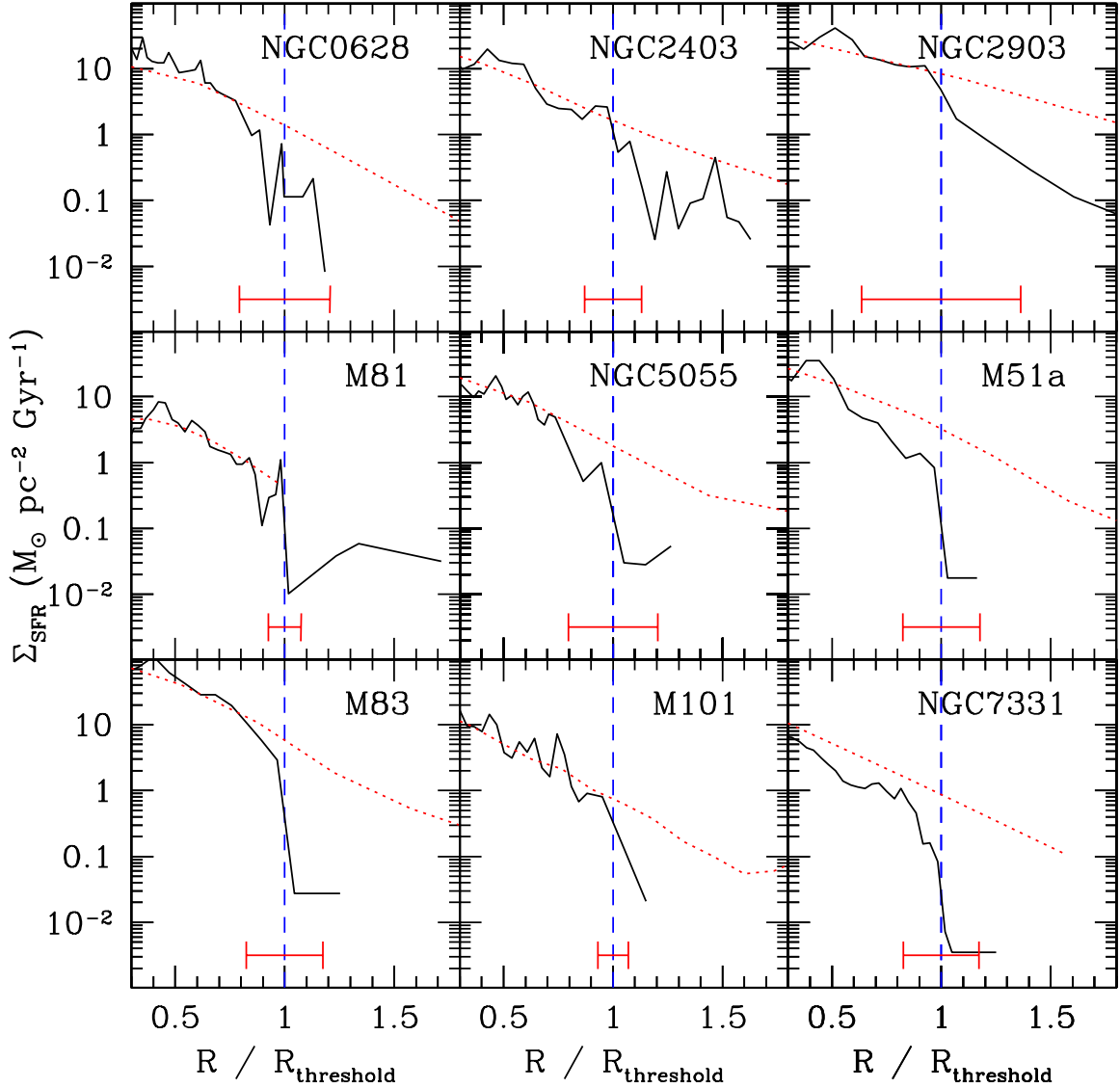}
\caption{\label{figthreshold1}Profiles of star formation rate surface density 
determined from the UV and corrected for extinction with the far infrared (dotted)
and from H$\alpha$ (solid, taken from Martin and Kennicutt, 2001) 
for the 9 galaxies in common with Martin and Kennicutt (2001),
having a threshold radius larger than 90 arcsec (so that our data
resolves it).
The radius is normalised by the ``threshold radius''
$R_{threshold}$ measured by Martin and Kennicutt (2001). We show only
a region around the threshold radius, marked by a vertical dashed
line. 
Note that for the H$\alpha$ profiles, the horizontal parts of the
profiles beyond $R/R_{threshold} \sim$ 1 are actually upper
limits. The H$\alpha$ profiles have a much higher resolution (10
arcsecs) than the UV + far infrared data (resolution shown by the
errorbar in the central-bottom part of each panel), however we have
many UV points beyond the threshold without any sign of a decrease in
the UV for most galaxies.  }
\end{figure*}

The above-described observation of a Schmidt-like behavior at very low
gas densities as derived from UV radial profiles suggests that
low-levels of star formation are in fact common beyond the usual
threshold radius, where H$\alpha$ emission ceases or is not (as
easily) observed.
We can verify that this is actually the case by taking the 9 galaxies
for which we have UV and infrared data, H$\alpha$ profiles from \citet{martin01}, and
for which the threshold radius given in \citet{martin01} is larger than 90 arcsec. 
This last condition ensures that we would resolve the threshold at our low
resolution. 
%
The star formation rate profiles derived from the UV and the H$\alpha$ 
emissions are presented in Figure \ref{figthreshold1}. Within the threshold
radius $R_{threshold}$, they are in rough agreement, while a systematic difference
occurs beyond it: eight out of the nine galaxies show no truncation (or 
any change of slope) in the UV profiles at the position of the H$\alpha$ 
threshold. The last one is M81, for which our UV profiles ends at the position of
the threshold. Note that we are limited for this galaxy by its large angular size: the
south outer part of the galaxy is getting close to the limit of the GALEX field of view.
The infrared background is also relatively bad so that we cannot extend the profiles
very far out in the disk. However, an inspection of the FUV image (at full resolution)
clearly reveals two spiral arms extending in the area between $R/R_{threshold}$ equals
1 and 1.5.

The apparent absence of threshold may seem at odds with the figures
presented in the appendix, in which an abrupt drop of the UV surface
brightness (or SFR surface density) is visible for a few
galaxies. This is the case for instance of NGC0891. However, for this
galaxy, the UV drop is concomitant with a drop in the gas density, and
occurs at quite large radius (35 kpc). Various reasons may affect the
gas at large radii (e.g. gravitational interactions, ram-pressure),
and obviously no star formation would occur when the gas reservoir was
removed. Thus, while abrupt drops exist in the UV for some galaxies,
they do not always correspond to a ``threshold'' of star
formation. NGC0925 presents a more interesting case where no UV is
detected beyond about 20 kpc, while HI is still observed. Even if this
indicates for this galaxy a ``UV threshold'', 
our analysis includes galaxies with detected star formation at similar
gas densities. Detailed studies of individual galaxies with high
resolution may help to understand the peculiarities of system like
this. This kind of analysis is beyond the scope of this paper.
Despite these questions on possible ``UV thresholds'', it is clear
from figure  \ref{figthreshold1} that beyond the usual threshold radius,
we frequently see  UV radiation when  H$\alpha$ ceased to be steadily found.
While many reasons for this are possible, including a variation in
the initial mass function, the simplest explanation is that at some
radius, star formation levels are so low that the number of high-mass
(short-lived) ionizing stars visible at any time becomes vanishingly
small, while slightly longer-lived stars, emitting in the UV for a
correspondingly longer time, are more frequent.  If this
interpretation is correct, the origin of the threshold of star
formation, largely debated on theoretical grounds
\citep{toomre64,quirk72,wang94,martin01,wongblitz,boissier03,joop04}
on the basis of H$\alpha$ observations would be simply the lack of
ionizing stars. 

\begin{figure}
\includegraphics[angle=0,scale=.40]{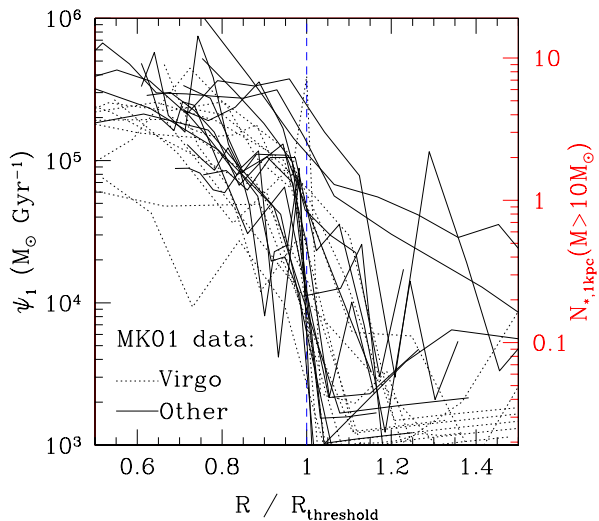}
\caption{\label{figthreshold2}
Profiles of the star formation rate surface density determined from
the H$\alpha$ profiles of \citet{martin01} (MK01). The left axis
indicates the star formation rate integrated in 1 kpc annular
ellipses. The right axis shows the number ionizing stars within the
same annular ellipses (see text for details). Dotted lines correspond
to galaxies in Virgo. The radius is normalised by the ``threshold
radius'' $R_{threshold}$ measured by \citet{martin01}. We show only a
region around the threshold radius, marked by a vertical dashed line.
Note that several points beyond $R/R_{threshold}$ (straight almost horizontal
lines) are actually upper limits.
}
\end{figure}

To test further this idea, we took the H$\alpha$ profiles of
\citet{martin01} around the threshold radius they defined. 
Since profiles are usually computed within annular ellipses, we
converted them to star formation rate integrated whithin 1 kpc 
wide annular ellipses (that we will call $\psi_1$), what is shown
in figure  \ref{figthreshold2}.
For galaxies at 17 Mpc (distance of the Virgo galaxies, dotted
lines), the resolution of 10 arcsec of \citet{martin01} correspond to 0.8
kpc, thus $\psi_1$ is actually close to the quantity really involved
in measurements.

This star formation rate $\psi_1$ at the threshold curiously shows
very low scatter around 4 10$^4$ M$_{\odot}$ Gyr$^{-1}$.
The right axis of Fig. \ref{figthreshold2} shows $\psi_1$ in the unit of
number of massive stars present at any time within the 1 kpc wide 
annular ellipse. To compute this number, we 
integrated the initial mass  function (assumed to be a 
Salpeter function over the range 0.1  and 100 $\rm M_{\odot}$) 
above 10 $\rm M_{\odot}$, and assumed a 
lifetime for these stars (10$^7$
yr). Our adopted limit of 10 $\rm M_{\odot}$ being conservative, 
our estimate should be an upper limit.
Note that the prediction of the number of O stars using {\it Starburst 99}
\citep{sb99} with Solar metallicity, and a Salpeter initial mass
function between 1 and 100 $\rm M_{\odot}$ is very close to the
numbers we obtained in that way. Introducing lower mass stars into the
initial mass function would reduce this number; thus once again, our
number is probably an upper limit to the number of ionizing stars.

From this Figure \ref{figthreshold2}, we conclude that
the threshold of star formation found by \citet{martin01} corresponds to
the radius where the number of ionizing stars at any given time within
1 kpc wide ellipses becomes lower than about 1. At the position
of the threshold, we would expect to find more than one star in
only about 19 \% of the galaxies of the galaxies of \citet{martin01}.
For three stars, the percentage drops to about 6 \%.
It is thus natural
that this radius will be the one where we stop finding HII regions,
not because of the absence of star formation, but because there is a
diminishingly small chance of catching a massive star in our annuli,
even with a normal initial mass function.
%

Although the threshold radius is defined by an abrupt drop in the
H$\alpha$ surface brightness profile, Martin \& Kennicutt (2001)
themselves noticed that many galaxies exhibit a few HII regions well
beyond this radius, where star formation does seem more stochastic
\citep[see also the HII regions found in extreme
outer parts of disk galaxies by][]{ferguson98}.
The presence of resolved young blue B stars in the outskirts
of M31 \citep{cuillandre01} also shows that some star formation
occurs below the critical gas surface density. Similarly,
\citet{davidge03} has shown that intermediate age stars
are presents at large radii in M33 and NGC2403, suggesting again that
star formation has proceeded in the past beyond the current threshold
radius. All these facts are totally consistent with our hypothesis
that the threshold radius is only marking the last radius where HII
regions are found in sufficient numbers to compute an H$\alpha$
profile, and not the end of star formation.

This is certainly relevant to the phenomenon of extended UV
(XUV hereafter) disks in which UV emission is observed at large radii (and
associated with relatively recent star formation) but where H$\alpha$
detections are rare \citep{gildepaz05,thilker05}.
The situation is however certainly not that simple since XUV disks are
themselves not obvious extrapolations of the inner UV profiles.  It
will, of course, be interesting to compare our results with the
relation between gas and star formation rate in XUV galaxies. We plan
to start that analysis with the spectacular case of NGC~4625 (Gil de
Paz et al., in preparation).

\section{Conclusion}

Using the galaxies of large angular size culled from the GALEX Atlas
of Nearby Galaxies \citep{gil06}, we have compiled GALEX UV and IRAS
IR images to compute profiles of attenuation and star formation
in 43 galaxies.
We find typically that UV attenuation decreases, from a few magnitudes
of extinction in the centers of galaxies, to low values at larger
radii. Combined with the metallicity gradients observed in those same
galaxies, this gives rise to a attenuation-metallicity relationship,
quantified in section \ref{secAz}.
No clear correlation is found between the attenuation and the gas surface
density.

Using the attenuation to correct the FUV profiles for dust, we derive
star formation rate profiles with which we test several modern
expressions of the classical ``Schmidt law''. Our data overlap with
those of \citet{kenni98a} but extend to significantly lower gas
densities suggesting that the UV light in general traces and reveals
lower levels of star formation than H$\alpha$, in coincidence with the
recent GALEX discoveries of UV extended (XUV) disks
\citep[e.g.][]{gildepaz05,thilker05}.  We suggest that the
much-debated threshold of star formation might have been an
observational selection effect caused by the extremely low number of
ionizing stars expected to be found beyond the ``threshold'' radius.
Indeed, for low levels of star formation, the H$\alpha$ radiation in
one elliptical annuli seems stochastic when the number of massive
stars in this annuli becomes close to $\sim$ 1. The UV radiation
coming from less massive and more long-lived stars (thus in larger
number) does not present a break in its radial profile at the
H$\alpha$ threshold radius.

While these results are already interesting in their own right, it is
obviously time to look to more galaxies, using Spitzer data in
combination with GALEX. Spitzer will allow us to include many more
galaxies in such studies owing to its better resolution than
IRAS. Indeed, a joint analysis of GALEX and Spitzer data is already
underway in a SINGS-GALEX collaboration. Note that the SINGS sample
will only partially overlap with the one presented here as some of the
closest galaxies were not included in SINGS.

\acknowledgments

GALEX (Galaxy Evolution Explorer) is a NASA Small Explorer, launched
in April 2003.  We gratefully acknowledge NASA's support for
construction, operation, and science analysis for the GALEX mission,
developed in cooperation with the Centre National d'\'Etudes Spatiales
of France and the Korean Ministry of Science and Technology. AGdP is
partially financed by the Spanish Programa Nacional de Astronom\'{\i}a
y Astrof\'{\i}sica under grant AYA2003-01676. We also thank the 
MAGPOP network for its support, and the referee for very constructive
comments.

{\it Facilities:} \facility{GALEX}

\appendix

\section{Appendix: individual profiles}

\begin{figure}
\includegraphics[angle=-90,scale=0.55]{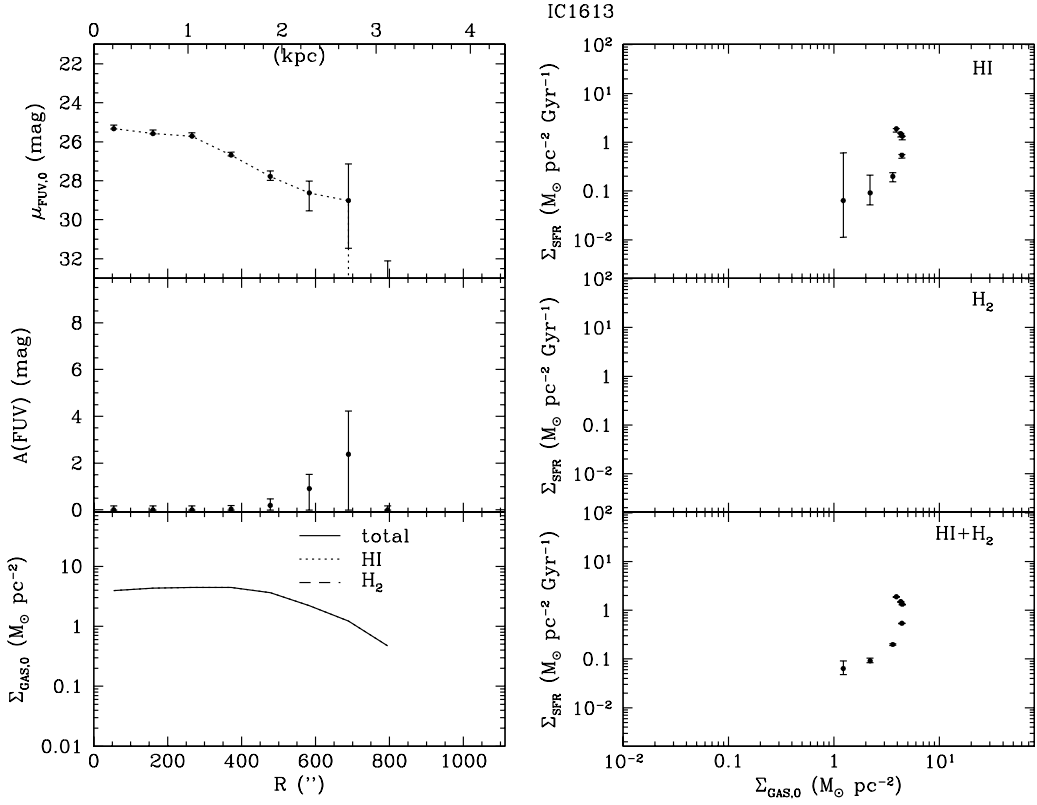}
\caption{\label{fprof1} [THIS IS ONE PANEL OF FIG. 9. THE COMPLETE FIG. 9
WILL BE AVAILABLE IN ELECTRONIC FORM ON THE APJ WEB SITE AND AT:
http://www.oamp.fr/boissier/preprint/preprint.html
]
Radial profiles of the gas surface density (bottom-left), FUV attenuation (middle-left),
FUV surface brightness (corrected for attenuation and to a face-on
inclination, top-left) when available. This FUV surface brightness is directly 
proportional to the star formation rate surface density shown in the right panels as a function of 
the HI, H2 and the total gasesous Hydrogen surface density along those
profiles. Errorbars in the left panels indicate how much the results would change
if our sky determination in the UV or far infrared was
moved by $\pm 1 \sigma$ (see section \ref{secerrorbar}).
The name of the galaxy is indicated at the top.}
\end{figure}

The left panels of Figure \ref{fprof1} shows the profiles of 
UV attenuation, UV surface brightness (corrected for attenuation and inclination) 
equivalent to a star formation rate surface density,
and of gas surface density used in this work.
The right panels show gaseous hydrogen surface densities (HI, H$_2$, HI+H$_2$)
vs the star formation rate surface density.

[included in electronic version]

\clearpage


\clearpage

\clearpage

\clearpage

\clearpage

\clearpage

\end{document}